\RequirePackage{fix-cm}
\documentclass[smallextended]{svjour3}       
\smartqed  
\usepackage{graphicx}

\usepackage[english]{babel}
\usepackage{color}
  \usepackage[all]{xy}
\usepackage{hyperref}
\usepackage{epsfig,amsfonts,amsmath,amssymb,latexsym}
\usepackage{multirow}

 \newcommand\be{\begin{equation}}
 \newcommand\ee{\end{equation}}
 \newcommand{\ba}{\begin{array}}
 \newcommand{\ea}{\end{array}}

\begin{document}


\title{Finite orbits of monodromies of rank two Fuchsian systems}

\author{Yuriy Tykhyy}



\institute{Y. Tykhyy \at
              Bogolyubov Institute for Theoretical Physics, 03143 Kyiv, Ukraine \\
              \email{tykhyy@bitp.kiev.ua}           
}

\maketitle


\begin{abstract}
We classified  finite orbits of monodromies of the Fuchsian system
for five $2\times 2$ matrices.  The explicit   proof of this
result is given.  We have proposed a conjecture for a similar
classification for $6$ or more $2\times 2$ matrices. Cases in
which all monodromy matrices have a common eigenvector are
excluded from the consideration. To classify the finite
monodromies of the Fuchsian system we combined two methods
developed in this paper. The first is an induction method: using
finite orbits of smaller number of monodromy matrices the method
allows the construction of such orbits for bigger numbers of
matrices. The second method is a formalism for representing the
tuple of monodromy matrices in a way that is invariant under
common conjugation way, this transforms the problem into a form
that allows one to work with rational numbers only.

 The classification developed in this paper can be considered as the first step
 to a classification of algebraic solutions of the Garnier system.
\end{abstract}


\section{Introduction}
Let us consider the Fuchsian system for $2\times2$ matrices:
\[Y(z) \in GL(2,{\mathbb C}): \frac{d}{d\,z} Y = \left({\sum_{k=1,}^{n}}_{a_k \neq \infty} \frac{A_k}{z - a_k}\right)Y.\]
Here $a_k$ are the branching points, i.e.  pairwise distinct numbers
on the Riemann sphere, and the following condition for the  matrices $A_k$ is implied:
$$\sum\limits_k A_k = 0.$$

Without the loss of generality we can put $ Tr\, A_k = 0, \forall
k$ and $Y(z) \in SL(2,{\mathbb C})$ for any $z$, and denote the
eigenvalues as follows: $eigen(A_k) = \pm \theta_k / 2$.

 We can perform an isomonodromic deformation for this system, i.e. move the
points $a_k$ simultaneously with such evolution of the matrices
$A_k$ that the monodromy of $Y$ around the branch points is
constant.
It gives us the Schlesinger system for $2\times2$ matrices, or the
Garnier system ${\mathfrak{ g}}_{n-3}$ (see
\cite{garn1},\cite{garn2}), where $a_k$ are independent variables,
and the elements of matrices $A_k$ are unknown functions. The
number of independent variables is $n-3$, because we can fix three
of the points $a_k$ as $0,\, 1,\,\infty$.

Now let us introduce the tuple of monodromy matrices. For this
purpose we introduce the collection of loops
$\gamma_1\,...\,\gamma_n$, as in the Fig.1. For each loop
$\gamma_k$ there is the monodromy matrix $M_k$. The product of all
monodromy matrices is equal to unity: $M_1\, M_2\,M_3\,...\,M_n =
{\mathbb I}$ and the eigenvalues of each monodromy matrix are
$eigen(M_k) = \exp(\pm i\pi \theta_k)$. Determinant of every
monodromy matrix equals $1$, due to the fact that the trace of
every matrix $A$ is zero.

  \begin{figure}[!h]
 \begin{center}
 $\;$\\ $\;$\\
 \resizebox{12cm}{!}{
\includegraphics{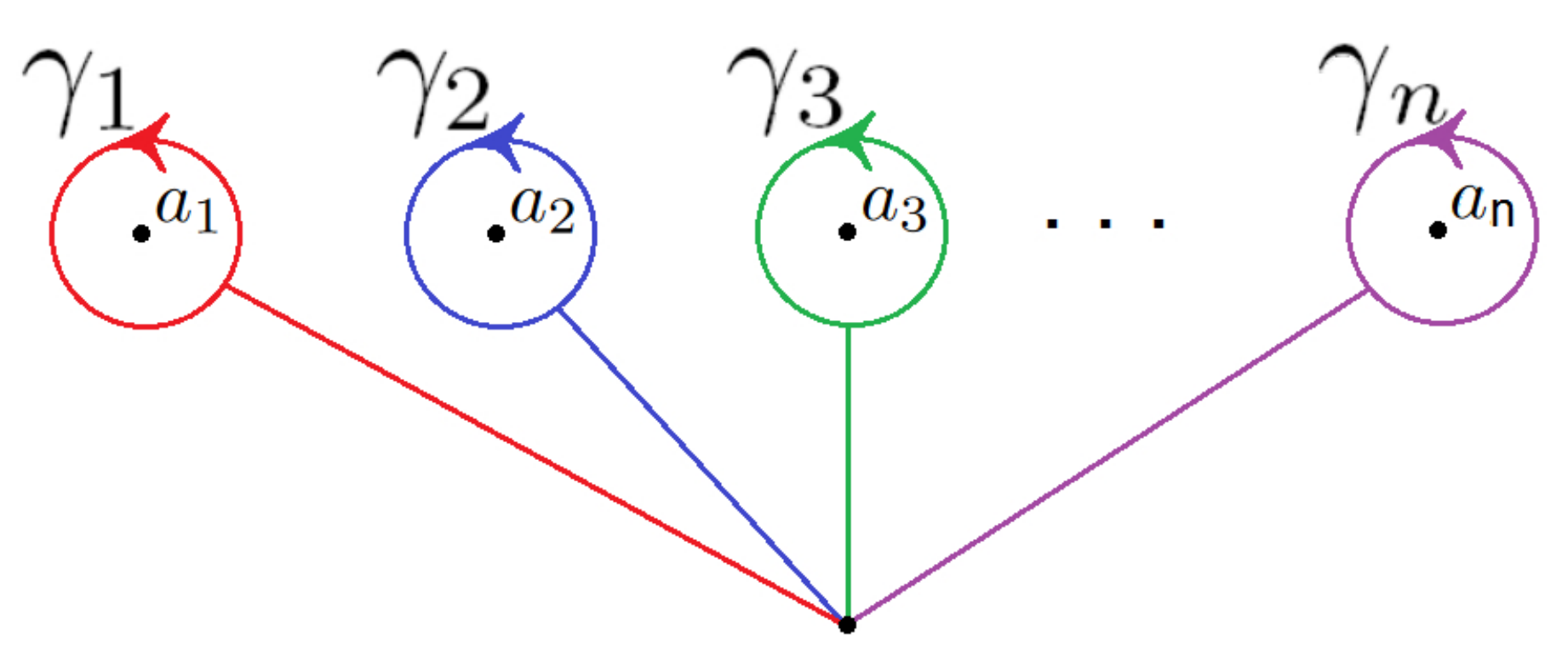}
 } \\  $\;$ \\ Fig. 1: Monodromy loops \\
 \end{center}
 \end{figure}

If the branching points move and interchange with each other, the loops braid and
 the monodromy matrices are transformed by an action of the braid group (see Fig. 2 and Fig. 3).
We will call this process ``braiding of matrices".

  \begin{figure}[!h]
 \begin{center}
 $\;$\\ $\;$\\
 \resizebox{8cm}{!}{
 \includegraphics{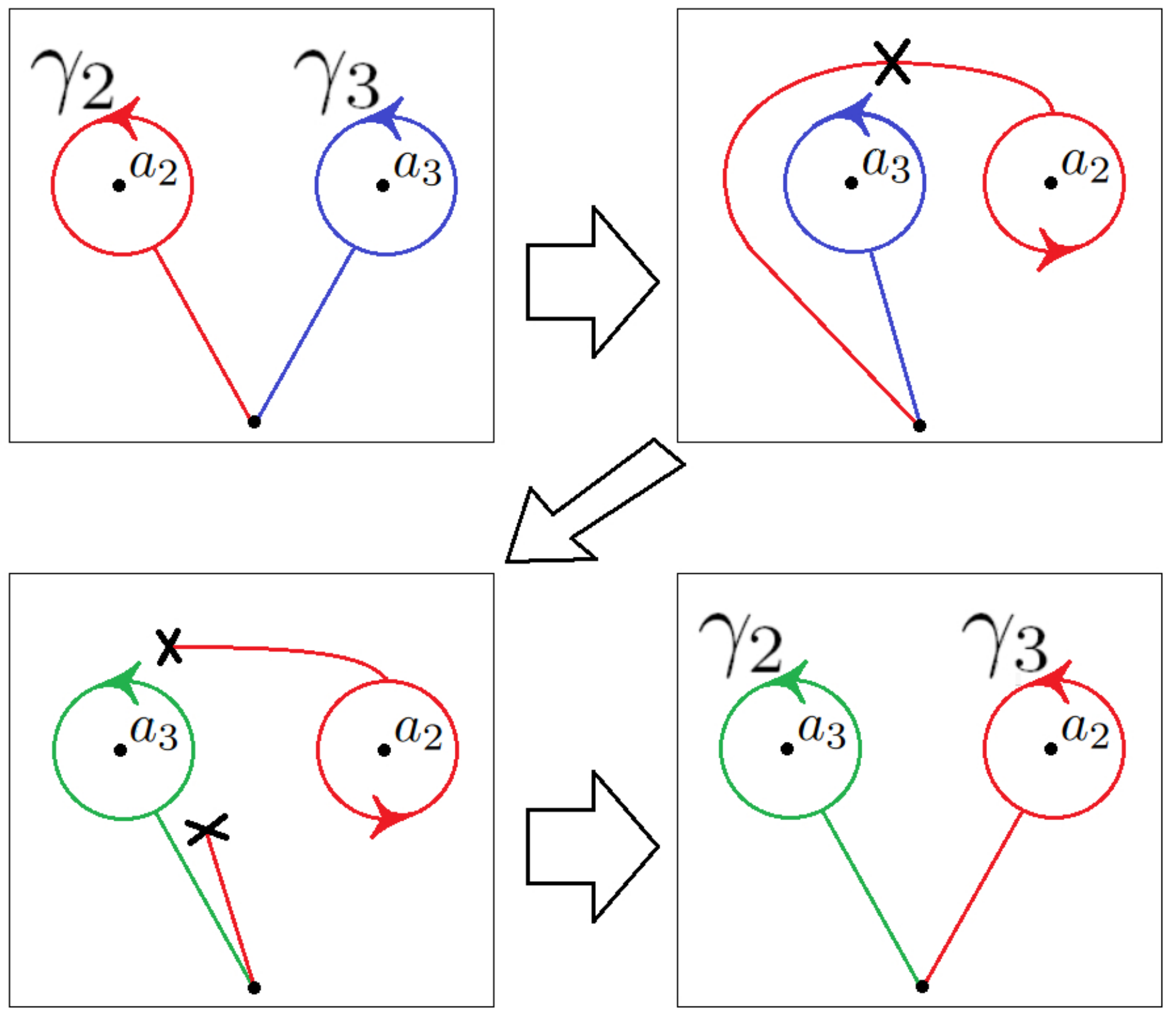}
 } \\  $\;$ \\ Fig. 2: Braid group action ${\mathcal B}_{2 3}$\\
 $\{M_2,\,M_3\} \mapsto \{M_2 M_3 M_2^{-1},\,M_2\}$\\
 When the branching points interchange, the corresponding loops braid.\\
 \end{center}
 \end{figure}

  \begin{figure}[!h]
 \begin{center}
 $\;$\\ $\;$\\
 \resizebox{8cm}{!}{
 \includegraphics{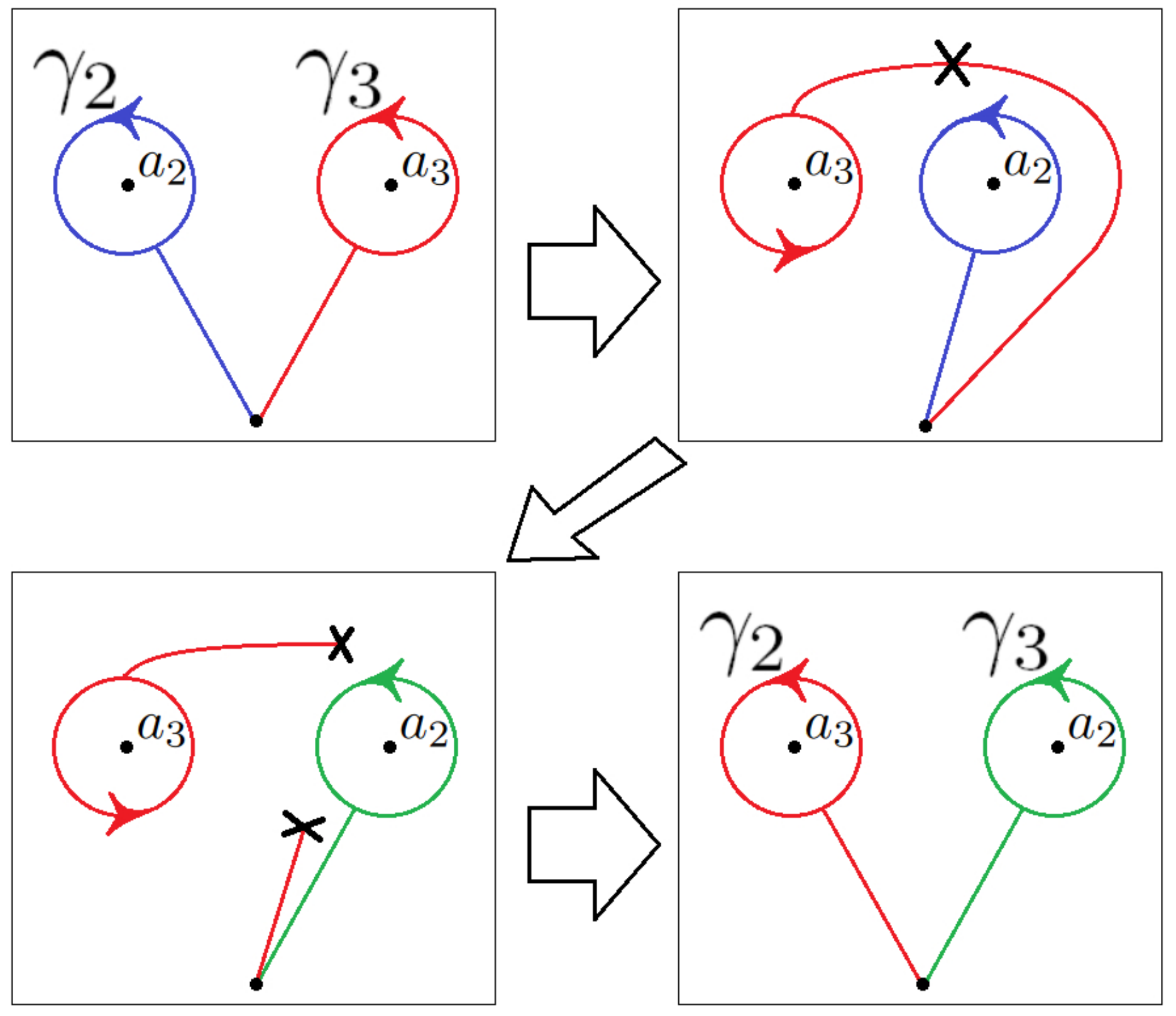}
 } \\  $\;$ \\ Fig. 3: Braid group action ${\mathcal B}_{3 2}$\\
 $\{M_2,\,M_3\} \mapsto \{M_3,\,M_3^{-1} M_2 M_3\}$\\
 \end{center}
 \end{figure}

The global problem is to classify the algebraic solutions of this
system. In this paper we solve a related problem: we classify the
finite monodromies of the Fuchsian system which will be a step
towards classification of algebraic solutions.

\textbf{Def:} We call the {\it finite monodromy} a tuple of monodromy matrices that
generates only a finite orbit under braid group action up to a common conjugation of matrices. $\square$

In order to define the braid group actions accurately let us
introduce some rules and notations. We declare that the case when
all the branching points $a_k$ have real positive values is
canonical. In this case the loops $\gamma_k$ are numbered from
left to right. If the branching points $a_k$ are not all real
positive numbers, then the loops corresponding to them are
numbered in order of increasing absolute values $|a_k|$, where the
infinity is considered to be the biggest in absolute value. In the
case of equal   absolute values the corresponding loops are
numbered in order of increasing $Arg(a_k)$. Here $Arg$ is the
argument of the complex number lying in the interval $(-\pi,\pi]$.

The domain of definition of the Garnier equation is the universal
covering of the space of the parameters $a_1\,...\,a_n$, which is $({\mathbb{C
P}}^1)^n$, excluding the cases when any two of the parameters $a_k$, $a_l$ coincide.

\textbf{Def:} We introduce the term {\it subbranch}. The universal
covering of the space of distinct $a_1,\,...\,a_n\,\, \in
\,\,({\mathbb{C P}}^1)^n$ can be divided into $n!$ parts, labelled
by members of the symmetric group $S_n$ in the following manner:
each collection of values $a_1,\,...\,a_n$ can be sorted as
described above, and the permutation of order of indices
correspond to the element $s$ of symmetric group. The exact
condition that a point of the universal covering belongs to the
part labelled by $s$ is

\[s\in S_n:\quad s(k) < s(l) \leftrightarrow\]
\[ \,|a_k| < |a_l|\,\, OR\,\, a_l = \infty\,\, OR\,\, \left(|a_k| = |a_l|\,\, AND
\,\, Arg(a_k) < Arg(a_l)\right).\]

Each such part is a disconnected space, and we call the connected
components of these parts {\it subbranches}. $\square$



For every subbranch we can introduce the object ${\mathcal
M}^{(n)}$, which is an element of the moduli space of the
monodromy.

\textbf {Def:} The object ${\mathcal M}^{(n)}$ is a tuple of $n$
matrices and $n$ integer values:
\[{\mathcal M}^{(n)}:\quad
\left\{M_1,\,M_2\,...M_n\,;\,N_1,\,N_2\,...\, N_n\right\},\]
where $M_1\,...\,M_n$ are $SL(2,{\mathbb C})$ monodromy matrices,
defined up to a common conjugation, the product of all
of them is the unity matrix and $N_1\,...N_n$ are
distinct integer numbers belonging to the interval $[1,n]$. Consequently, the object
${\mathcal M}^{(n)}$ is a member of the following set:
\[{\mathcal M}^{(n)} \in SL(2,{\mathbb C})^{n-1} /
SL(2,{\mathbb C})\,\times\,S_n.\]

Here each matrix $M_k$ is the monodromy matrix corresponding to
the loop $\gamma_k$, and $N_k$ is an integer defined as follows:
if we denote the branching point corresponding to the loop
$\gamma_k$, as $a_m$, then $N_k = m$. And if the subbranch which
this tuple of monodromy matrices corresponds to, is labeled by the
element $s$ of the symmetry group, then $\gamma_k$ is the loop
around the point $a_{s(k)}$, and $N_k = s(k)$.

We call the form of  ${\mathcal M}^{(n)}$ with $N$-values its {\it
long form}, and its form  without $N$-values its {\it short form}.
It will be enough to consider the short form only in the majority
of cases treated in the present paper. $\square$

Therefore,  the values $N_1...N_n$ are constant in every
subbranch, and the monodromy matrices are constant there up to a
common conjugation.

Next, let us define the rule of moving from one subbranch to another.

If  two neighboring branching points interchange, then loops
corresponding to these points braid and should be redefined to
recover the normal form. In this case  two corresponding values
$N$ interchange, and two matrices $M$ are transformed by braid
group action. The braid group action interchanges two matrices and
conjugates one of them with another one.

We have two types of such the actions, each interchanges two neighboring
branching points, as illustrated in Fig. 2 and Fig. 3 respectively:
\[{\mathcal B}_{k, k+1}:\]
\[\left\{\ba{cccccccc}M_1,&...&M_{k-1},&{\color{red}M_{k}},&{\color{blue}M_{k+1}},& M_{k+2},&...&M_n;\\
N_1,&...&N_{k-1},&{\color{red}N_{k}},&{\color{blue}N_{k+1}},&N_{k+2},&...&N_n\ea\right\}\rightarrow\]
\[\left\{\ba{cccccccc}M_1,&...&M_{k-1},&{\color{blue}M_k M_{k+1}M_k^{-1}},&{\color{red}M_k},& M_{k+2},&...&M_n;\\
N_1,&...&N_{k-1},&{\color{blue}N_{k+1}},&{\color{red}N_k},&N_{k+2},&...&N_n\ea\right\},\]

\[{\mathcal B}_{k+1, k}:\]
\[\left\{\ba{cccccccc}M_1,&...&M_{k-1},&{\color{red}M_{k}},&{\color{blue}M_{k+1}},& M_{k+2},&...&M_n;
\\N_1,&...&N_{k-1},&{\color{red}N_{k}},&{\color{blue}N_{k+1}},&N_{k+2},&...& N_n\ea\right\}\rightarrow\]
\[\left\{\ba{cccccccc}M_1,&...&M_{k-1},&{\color{blue}M_{k+1}},&{\color{red}M_{k+1}^{-1}M_k M_{k+1}},& M_{k+2},&...&M_n;\\
N_1,&...&N_{k-1},&{\color{blue}N_{k+1}},&{\color{red}N_k},&N_{k+2},&...&N_n\ea\right\},\]
or briefly
\be\label{braida1}{\mathcal B}_{k, k+1}: \quad M_k \rightarrow M_k
M_{k+1} M_k^{-1},\quad M_{k+1} \rightarrow M_k,\quad N_k
\leftrightarrow N_{k+1},\ee

\be\label{braida2}{\mathcal B}_{k+1, k}: \quad M_k \rightarrow
M_{k+1},\quad M_{k+1} \rightarrow M_{k+1}^{-1} M_k M_{k+1},\quad
N_k \leftrightarrow N_{k+1}.\ee The two indices of ${\mathcal B}$
are the numbers of braided loops and must differ by  $\pm 1$.
Informally we will call the braid group actions {\it braiding}.

It should be noticed that all features of the tuple of monodromy
matrices are symmetric under cyclic permutation of matrices and
$N$-values, hence index $k$ of  $M_k$ matrix may be treated as an
integer modulo $n$. Thus, in total we have $2\,n$ braid group
actions.

Armed with these definitions, let us proceed to formulating the
problem of finite orbits of the tuples of monodromy matrices under
braid group actions.

If the solution of the Garnier system is algebraic, then  it has a
finite branching. Thus in order to classify the algebraic
solutions we have to classify the finite orbits of the braid group
acting on monodromies.

The goal of this paper is a classification of all finite orbits
of the braid group action on monodromies.

\textbf{Def:} We call the ${\mathcal M}^{(n)}$ {\it triangular} if all matrices have a common eigenvector, therefore
can be made simultaneously to be lower-triangular. $\square$

\textbf{Def:} The {\it orbit} is the set of all ${\mathcal M}^{(n)}$'s, obtained from
one of them, by the  action  of the braid group. $\square$

We denote the ${\mathcal M}^{(n)}$ which belongs to a finite orbit with a subscript $F$: ${\mathcal M_F}^{(n)}$. Note that ${\mathcal M}^{(n)}$ and
${\mathcal M_F}^{(n)}$ refer to fixed tuples of matrices, not to
sets of all such tuples. $\square$

\textbf{Def:} We will call the operation of replacing of two
neighboring matrices by their product (using the short form) the
{\it reduction} from a ${\mathcal M}^{(n)}$ to ${\mathcal
M}^{(n-1)}$. $\square$

\textbf{Def:} The inverse operation is the following: replacing
any matrix by two matrices where one of them is an arbitrary
$SL(2,{\mathbb C})$ matrix and another one is such that their
product equals the original matrix.
We will call this to be an {\it induction} from a ${\mathcal
M}^{(n)}$ to ${\mathcal M}^{(n+1)}$. A special type of induction
is an {\it addition of the unit matrix}. $\square$

The {\it length} of the orbit is defined as the number of branches
in the transformation of monodromy. In terms of ${\mathcal
M}^{(n)}$ the length is the number of  such members of the orbit
that their $N_k$'s have trivial permutation: $N_k = k, \ \forall
k$. Therefore, the total number of members of the orbit is
$length\times n!$. This definition of the length is the only
reason to define ${\mathcal M}^{(n)}$ to be not simply a tuple of
matrices, but a tuple of matrices associated with a tuple of
numbers. The number of members of the orbit with different tuples
of matrices (not taking into account the permutations) is the
number which is a multiple of the length and a divisor of
$length\times\,n!$. Nevertheless, we do not  define the length by
this number, because such the definition would be mathematically
unnatural.

Our problem is classification of all finite orbits. In this paper
the problem has been solved for ${\mathcal M}^{(5)}$, excluding
the {\it triangular} case.

It is important to note that there are some symmetries for
${\mathcal M}^{(n)}$ which are not equivalences: the cyclical
permutation of all monodromy matrices (see (\ref{symgroup1}));
multiplying any two matrices by $-1$ (see (\ref{symgroup2}));
taking inverses of all matrices, simultaneously with reversing
their order (see (\ref{symgroup3})); complex conjugation of all
elements of all matrices (this symmetry will not be used in the
paper):

\be\label{symgroup1}M_1,\,M_2\,...\,M_n\rightarrow
M_{2},\,M_{3}\,...\,M_{n},\,M_1;\ee \be\label{symgroup2}
M_1,\,...M_p,\,...M_q,\,...\,M_n\rightarrow
M_1,\,...-\!M_p,\,...-\!M_q,\,...\,M_n;\ee \be\label{symgroup3}
M_1,\,M_2\,...\,M_n\rightarrow
M_n^{-1},\,M_{n-1}^{-1}\,...\,M_1^{-1}.\ee

Note that operation (\ref{symgroup1}) can be represented as a
series of braidings ${\mathcal B}_{1,2}$, ${\mathcal
B}_{2,3}$,...${\mathcal B}_{n-1,n}$. That is why every orbit is
closed under this operation.

For $n\leq 3$ the problem formulated above is trivial: any
${\mathcal M}^{(1)}$, ${\mathcal M}^{(2)}$ and ${\mathcal
M}^{(3)}$ belong to the orbit of the length $1$. Also, there
exists a trivial method of transforming  of ${\mathcal M_F}^{(n)}$
into ${\mathcal M}^{(n+1)}$ (which also belongs to a finite
orbit): this is achieved by addition of the unit matrix. Note that
for any $n$ there exists a trivial case when all monodromy
matrices commute.

The problem posed above is solved for the case of $n=4$ in
\cite{painleve_6}. That solution is used in the present paper as a
base for the construction finite orbits of ${\mathcal M}^{(5)}$'s.

{\bf Acknowledgements.} The research described in this paper was
supported by the National Academy of Sciences of Ukraine (project
No. 0117U000238). The author would like to thank Dr. O.Lisovyy for
formulation of the problem, and Dr. N.Iorgov for fruitful
discussions.

\section{Signature formalism}

Now we choose the most convenient formalism for classification of
${\mathcal M_F}^{(n)}$'s - elements of the moduli space of the
monodromy, which belong to finite orbits.

First, we note that it is better to notate ${\mathcal
M_F}^{(n)}$'s not by elements of the matrices, but by traces of
different products of these matrices. This is because the common
conjugation, which is by definition a trivial transformation at
the moduli space of the monodromy, acts trivially on these traces,
but not on the elements of matrices.

Further, since we are looking for ${\mathcal M}^{(n)}$'s belonging
to finite orbits, the products of their matrices are likely to be
roots of unity, so their traces are likely to have a form

\[2\,\cos(\pi {\mathbb Q}),\]

where ${\mathbb Q}$ means the set of rational numbers. Since for
rigorous classification it is better to operate with rational
numbers, we propose a described below formalism.

\textbf {Def:} {\it Signature}. For each ${\mathcal M}^{(n)}$ a
special collection of values can be calculated. It is called
the {\it signature}.

The {\it signature} consists of  sub-collections of the following values:
\\
1) the $\theta$ value for any matrix:

\be\label{thetas}\forall x \in [1,n]:\,\, \theta_x = \frac{1}{\pi}
\arccos\frac{1}{2} Tr\left(M_{x}\right),\ee

where the eigenvalues of  $M_x$ are $\exp(i\pi \theta_x)$ and
$\exp(-i\pi\theta_x)$.
\\
2) the $\sigma$ value for any subsequence of two or more neighboring
matrices in ${\mathcal M}^{(n)}$

\be\label{sigmas1}\forall x \in [1,n] \,\,\forall y \in [x+1,
x+n-2]:\,\, \sigma_{x,\,x+1\,...y} = \frac{1}{\pi}
\arccos\frac{1}{2}Tr\left(\prod_{z = x}^{y} M_{z\, mod\,
n}\right),\ee

where the product $\prod_{z = x}^{y} M_{z\, mod\, n}$ has
eigenvalues $\exp(i\pi\sigma_{x...y})$ and $\,\,\,\,\,\,\,\,$
$\exp(-i\pi\sigma_{x...y})$.

3) the $\sigma$ value for any two not intersecting subsequences of
neighboring matrices is the following one:

\[\forall x \in [1,n] \,\,\forall y \in [x, x+n-4] \,\,\forall p \in [y+2, x+n-2] \,\,\forall q \in [p,
x+n-2]
 :\]
 \be\label{sigmas2}\sigma_{x,\,x+1,\,...y,\,p,\,p+1,\,...q} = \frac{1}{\pi} \arccos\frac{1}{2}Tr\left(
\prod_{z = x}^{y} M_{z\, mod\, n}\prod_{z = p}^q M_{z\, mod\,
n}\right).\ee

Anyway, every $\theta$ or $\sigma$ depends on the trace of a product of matrices, and the tuple of indices of $\theta$ or $\sigma$ means the indices of these matrices.

Due to the definition, all $\theta$'s and $\sigma$'s are
determined  modulo $2$ and up to a sign. The order of indices in
$\sigma$ is important, but the cyclical permutation of the indices
is treated as an equivalence:

\[\sigma_{a,b,...c} \equiv \sigma_{b,...c,a};\quad \sigma_{a,c,b} \neq \sigma_{a,b,c}.\]

 The following notations are also treated as equivalent:

\be\label{eqsigmas} \theta_{x} \equiv \sigma_{x+1,...x+n-1},\quad
\sigma_{x,x+1,...y} \equiv \sigma_{y+1,y+2,...x+n-1}.\ee

Indeed, $M_x\cdot M_{x+1}\cdot...\cdot M_y =
\left(M_{y+1}\cdot...\cdot M_{x+n-1}\right)^{-1}$, so the corresponding traces are  equal. $\square$

 We call the signature of ${\mathcal M}^{(n)}$ to be $n$-signature. The
reason of using such formalism is that  $\theta$ and $\sigma$
values for a ${\mathcal M_F}$ are almost always rational numbers.

We call the signature {\it inconsistent}, if no tuple of matrices
which corresponds to it exists.

Each allowed tuple of indices in the signature we call the {\it
cell}. If there is one index (corresponding to a trace of one
monodromy matrix) -- the cell is called $\theta$, if there are
more indices than one -- the cell is  called $\sigma$.  We say
that two tuples of indices  $(x,\,x+1,\,...y)$,
$(y+1,\,y+2,\,...x+n-1)$ or $x$, $(x+1,\,...x+n-1)$ according to
(\ref{eqsigmas}) define the same cell.

For example in the case of  ${\mathcal M}^{(4)}$, denoted as
\[M_1,\, M_2,\, M_3,\, M_4,\,\, M_1 M_2 M_3 M_4 = {\mathbb I}\]
the signature consists of eight numbers:
\[\theta_1,\,\theta_2,\,\theta_3,\,\theta_4,\,\sigma_{12},\,\sigma_{23},\,\sigma_{13},\,\sigma_{24}.\]

In the case of $n=5$ the signature consists of 20 cells, in the
case of  $n=6$ it consists  of 39 cells, and in general for any
$n$ it consists of  $n(n-1)(n^2-5 n+12)/12$ cells: $n$ of them are
$\theta$'s and $n(n-3)(n^2 - 3 n + 8)/12$ of them are $\sigma$'s.

\textbf {Def:} We call by the {\it particular signature} a
signature in which some cells are undefined: no values for these
cells are defined. $\square$

\textbf {Def:} We call by the {\it incomplete signature} a
special case of the particular signature when for some indices
$a,b$ such that $b-a=\pm 1$ all the $\sigma$'s which contain index
$a$ and do not contain index $b$, are undefined, but the rest of cells are
defined. In case if there are two notations for one cell, (see
(\ref{eqsigmas})) and at least one of them does not contain $a$ or
contains $b$, then this cell must be defined. $\square$

In the incomplete signature all the $\theta$'s are defined,
including the $\theta_a$: although in the $\theta_a$ the index $a$
is present and the index $b$ is absent, but the $\theta_a$ can be
notated as $\sigma_{a+1,...a-1}$.

In the  $n = 5$ case the incomplete signature contains 16 cells.

The example of an incomplete signature for $n=4$ with $a = 1$ and
$b = 2$ is this: we  take all cells despite those which contain
index $1$ and do not contain  index $2$. The $\sigma_{12} =
\sigma_{34}$ is defined; the $\sigma_{14}$ seems to be undefined,
but it is equal to $\sigma_{23}$, which is defined. The
$\sigma_{13}$ is undefined, and the $\sigma_{24}$ is defined. In
total, this incomplete signature consists of seven values, and as
it is known, it is enough to reconstruct all matrices up to common
conjugation, except for the {\it triangular} case.

We call by  {\it merging} of two particular signatures a procedure
of making of them one signature or particular signature by filling
in the cells. It is impossible to merge two particular signatures
if they have at least one cell which is defined in both particular
signatures, but have different values. Value of the cell which is
defined in both merging particular signatures and coincide, or
defined only in one -- is retained; the cell which is undefined in
both signatures -- remains undefined.

We say that two signatures, incomplete signatures or particular
signatures {\it coincide} if in all cells, which are defined in
both of them, the values coincide.

{\lemma{$\,$}\label{lemma_signature}} The signature (and also the
incomplete signature) is sufficient for unique reconstruction of
${\mathcal M}^{(n)}$ (up to simultaneous conjugation), except for the {\it triangular} case
(where all matrices will have a common eigenvector, therefore can be
made simultaneously to be lower-triangular, because in this case the
$[2,1]$ elements of matrices can have arbitrary values and do not
affect the signature).

\textbf{Proof:} To generalize the proof for signature and
incomplete signature, we re-formulate the problem: there is a
linear tuple of matrices, we know the traces of every matrix, of
every product of subsequence of neighboring matrices, and of the
product of each two non-intersecting subsequences. For an
incomplete signature, in which there are undefined cells
containing index $a$ and not containing $b$, we consider a linear
tuple of all matrices except for $M_a$; then, after the
reconstruction of all other matrices, we will be able to
reconstruct $M_a$ using the fact that the product of all matrices
must make the unit matrix.

To prove the lemma, let us consider three cases:

\textbf{Case 1:} The general case. In this case, there exists at
least one pair of neighboring matrices which have no common
eigenvector. Let us call them $M_p$ and $M_q$. The non-existence
of a common eigenvector  can be checked using the condition
\be\label{nocommon} Tr(M_p)^2 + Tr(M_q)^2 + Tr(M_p\cdot M_q)^2
\neq Tr(M_p)Tr(M_q)Tr(M_p\cdot M_q) + 4, \ee taking into account
that both matrices have determinant $1$.

\textbf{Case 1a:} The most general case is the most simple one:
one of the above two matrices, let's $M_p$, has the trace not
equal to $\pm 2$.

The $M_p$ can be diagonalized. We make  simultaneous
conjugation of all the matrices such that $M_p$ turns out to be
diagonal with different diagonal elements.

Now we know all diagonal elements of all matrices: for any $r\neq
p$ we have $M_r[1,1] + M_r[2,2] = Tr(M_r)$ and $M_r[1,1]M_p[1,1] +
M_r[2,2]M_p[2,2] = Tr(M_r\, M_p)$. This is the system of two
equations for $M_r[1,1]$ and $M_r[2,2]$. The system has a unique
solution because the matrix elements $M_p[1,1]$ and $M_p[2,2]$ are
the eigenvalues of $M_p$ and  are distinct by our assumption.

Then, due to the condition that $M_p$ and $M_q$ have no common
eigenvectors, the non-diagonal matrix elements $M_q[1,2]$ and $M_q[2,1]$
are both nonzero, and we can perform such the conjugation that $M_p$ remains
diagonal and  $M_q[2,1]$ turns $1$.

Now we can reconstruct the non-diagonal elements of any other
monodromy matrix, let us call it $M$, using the non-degenerate
system of two linear equations:
\[M[1,2] M_q[2,1] + M[2,1] M_q[1,2] = Tr (M_q \cdot M)  - M_q[1,1] M[1,1] - M_q[2,2] M[2,2],\]
\[M[1,2] M_p[2,2]M_q[2,1] + M[2,1] M_p[1,1] M_q[1,2] = \]
\[ = Tr(M_p\cdot M_q\cdot M)   - M_p[1,1] M_q[1,1]M[1,1] -
M_p[2,2]M_q[2,2]M[2,2].\]

\textbf{Case 1b:} Similar calculations can be performed in the case when  $M_p$ and
$M_q$ both have traces equal to $\pm 2$:

Let us consider the case $Tr M_p = 2$. This matrix can't be unity,
because it has no common eigenvectors with $M_q$. That is why it
is possible, using the common conjugation, to put $M_p$ to lower
triangular form, with all non-zero elements equal to 1. Since
matrices have no common eigenvector then $M_q[1,2] \neq 0$.
Moreover, we can select such a conjugation that $M_q[1,1] = 0$,
and then $M_q[2,2] = \pm 2$ and  $M_q[2,1] = -1/M_q[1,2]$. Now we
can reconstruct all elements of any other matrix $M$ using the
system of equations:
\[M[1,2] = Tr(M\cdot M_p) - Tr(M),\]
\[M[2,2]M_q[1,2] = Tr(M\cdot M_p\cdot M_q) - Tr(M\cdot M_q),\]
and take $M[1,1]$ from $Tr M$, and the $M[2,1]$ -- from $Tr (M
\cdot M_q$).

\textbf{Case 2:} There exist two non-intersecting subsequences of
neighboring matrices $M_p,...,M_q$ and $M_s,...,M_t$ ($p\leq q <
s\leq t$), such that products $M_p\cdot...\cdot M_q$ and
$M_s\cdot...\cdot M_t$ have no common eigenvector.

Let us construct matrices $M_p\cdot...\cdot M_q$ and
$M_s\cdot...\cdot M_t$ similarly to the Case 1. For any $r<p$ we
reconstruct the elements of the matrix which is the product
$M_r\cdot...\cdot M_{p-1}$ similarly to Case 1, knowing the values
of the following traces: $Tr(M_r\cdot...\cdot M_{p-1})$,
$Tr(M_r\cdot...\cdot M_{p-1}\cdot M_p\cdot...\cdot M_q)$,
$Tr(M_r\cdot...\cdot M_{p-1}\cdot M_s\cdot...\cdot M_t)$,
$Tr(M_r\cdot...\cdot M_{p-1}\cdot M_p\cdot...\cdot M_q\cdot
M_s\cdot...\cdot M_t)$, exact form of the matrices
$M_p\cdot...\cdot M_q$ and $M_s\cdot...\cdot M_t$ and the
condition that they have no common eigenvector.

Hence we can reconstruct all matrices $M_r$ for $r<p$.

In similar manner we can reconstruct matrices $M_r$ for $q<r<s$
and $r>t$.

Finally, for every $r \in [p,q)$ we can reconstruct the matrix
$M_p\cdot...\cdot M_r$ using the known values of four traces: $Tr(M_p\cdot...\cdot
M_r)$, $Tr(M_p\cdot...\cdot M_r\cdot M_s\cdot...\cdot M_t)$,
\[Tr(M_p\cdot...\cdot M_r\cdot M_p\cdot...\cdot M_q) = Tr(M_p\cdot...\cdot M_r)Tr(M_p\cdot...\cdot M_q) - Tr(M_{r+1}\cdot...\cdot M_q)\]
and in a similar way we calculate the value of
$Tr(M_p\cdot...\cdot M_r\cdot M_p\cdot...\cdot M_q\cdot
M_s\cdot...\cdot M_t)$. This is the linear system of four
equations for four variables (four elements of the matrix
$(M_p\cdot...\cdot M_r)$) which has a unique solution due to the
fact that two matrices $(M_p\cdot...\cdot M_q)$ and
$(M_s\cdot...\cdot M_t)$ have no common eigenvector.

Therefore we know all matrices $M_r$ for $r\in [p,q]$ and, after the same procedure, for
$r\in [s,t]$.

\textbf{Case 3:}

Every two matrices and every two products of non-intersecting
subsequences of neighboring matrices have a common eigenvector.

This case can be {\it triangular} only: all matrices have a common eigenvector.

Indeed, assume that this assumption is wrong.

First, in this case a monodromy matrix with only one eigenvector
cannot exist because this eigenvector would be a common for all
the matrices.

Therefore, if there exist matrices with traces equal to $\pm 2$, then they
are proportional to unit matrix (are scalar matrices), and we can exclude them, retaining our
knowledge about traces, and equivalently considering the case without
these matrices. If all matrices are  scalar ones, then this contradicts the  assumption
 of non-existence of common eigenvector of all matrices, so we will assume that there
 exists at least one non scalar matrix.

Then there remain only matrices with traces different from $\pm
2$.

The matrix $M_1$ can be made diagonal. Its eigenvectors are $v_1$
and $v_2$ (see \ref{vectors}). For all other matrices one of these
vectors is an eigenvector, but there exists at least one matrix,
let us call it $M_p$, for which $v_1$ isn't  eigenvector (but
$v_2$ of course is), and at least one other matrix, let us call it
$M_q$, for which $v_2$ isn't eigenvector (but $v_1$ is).
Therefore, after excluding scalar matrices, the tuple contains at
least three matrices $M_1$, $M_p$ and $M_q$. The  matrices $M_p$
and $M_q$ must have a common eigenvector, let us call it $v_3$,
and, using the common conjugation not affecting $v_1$ and $v_2$,
we can transform this vector to the form (\ref{vectors}):
\be\label{vectors}v_1 = \left(\ba{c}1\\0\ea\right),\quad v_2 =
\left(\ba{c}0\\1\ea\right),\quad v_3 =
\left(\ba{c}1\\1\ea\right).\ee Therefore, any other matrix must
have a common eigenvector with $M_1$, $M_p$ and $M_q$, and that is
why the eigenvectors of every matrix must be two of the three
vectors $v_1$, $v_2$ and $v_3$.

It means that every matrix must belong to the one of three types
(\ref{matrtypes}), and for each type there exists at least one
matrix:
\[ M =
\left(\ba{cc}e^{i\pi\theta}&0\\0&e^{-i\pi\theta}\ea\right),\quad M
=
\left(\ba{cc}e^{i\pi\theta}&0\\e^{i\pi\theta}-e^{-i\pi\theta}&\quad
e^{-i\pi\theta}\quad \ea\right),\] \be\label{matrtypes} M =
\left(\ba{cc}\quad e^{i\pi\theta}\quad
&e^{-i\pi\theta}-e^{i\pi\theta}\\0&e^{-i\pi\theta}\ea\right).\ee
Then there must exist at least one pair of neighboring matrices
belonging to different types. Let it be $M_r$ with the
eigenvectors $v_1$ and $v_2$, and $M_{r+1}$ with the eigenvectors
$v_1$ and $v_3$. But their product does not belong to any of three
above types (to see this we take into account that traces of all
matrices differ from $\pm 2$). The matrix $(M_r\cdot M_{r+1})$ has
the eigenvector $v_1$, but not $v_2$ or $v_3$. By the way, there
exists at least one matrix of such type that its eigenvectors are
$v_2$ and $v_3$. This is $M_p$. It has no common eigenvectors with
$(M_r\cdot M_{r+1})$.

Therefore the above assumption is wrong and it means that the considered tuple is {\it triangular}:
all matrices in the considered tuple
have a common eigenvector.

Lemma \ref{lemma_signature} has been proven. $\square$

\textbf{Notice:}    There exist inconsistent signatures for which a  tuple
of monodromy matrices does not exist.

For every signature, or incomplete signature, we have three
possibilities:

1. tuple of matrices does not exist (the signature is inconsistent),

2.  there exists only one tuple of matrices and it has no common eigenvector
for all matrices (non-triangular case),

3. there exist many tuples of matrices,  such
that all matrices have common eigenvector and can be
simultaneously made lower-triangular (triangular case).

At a first glance,  the signature contains excessive information
about the collection of matrices ${\mathcal M}^{(n)}$. Indeed, due
to its definition the collection of matrices ${\mathcal M}^{(n)}$
has only $3 n-6$ degrees of freedom.  But the corresponding
signature contains $n(n-1)(n^2-5 n+12)/12$ cells.

But if we use smaller tuple of $\sigma$'s
 than defined in (\ref{sigmas1},\ref{sigmas2}) -- several discrete options for reconstruction of the
tuple of matrices remain. For example for the tuple  ${\mathcal M}^{(4)}$ we have $4\times
3-6=6$ degrees of freedom. But if we use only six cells of the
signature --- four $\theta$'s and two $\sigma$'s ---  two
options remain. Anyway any smaller tuple of $\sigma$'s  will be
not enough for the Lemma \ref{lemma_signature}.

This formalism is similar to formalism of $p$ values developed in \cite{mazocco}, and every $\sigma$
from the present formalism is
\[\sigma = \frac{1}{\pi}\arccos(\frac{p}{2}).\].

\section{The list of signatures of ${\mathcal M_F}^{(4)}$'s}

Here we present the list of 4-signatures which correspond to
${\mathcal M_F}^{(4)}$'s.

The present list is obtained by our computer program, and was
compared with list in the paper \cite{painleve_6} (Theorem 1 and
Table 4) to make sure that it is obtained correctly.

To shorten the list we present only one member of
each orbit.

The present list differs from the list of \cite{painleve_6} in the following
four aspects: in majority of orbits another element of the orbit is
presented; here we present  $\theta$'s, while two or three
different tuples of $\theta$'s can correspond the same tuple of
$\omega_x,\,\omega_y,\,\omega_z,\,\omega_4$. That is why the list
turned out to be almost three times longer; notations
$\sigma_{23},\,\sigma_{13},\,\sigma_{12}$ are used instead of
$r_x,\,r_y,\,r_z$; $\sigma_{24}$ is also presented.

Therefore, in this list, see Table 1 a signature with three
parameters is presented, which corresponds to any triangular tuple
(it can correspond to different orbits of ${\mathcal M}^{(4)}$'s);
orbit $2$ is obtained from the orbit of ${\mathcal M}^{(3)}$ with
three arbitrary parameters by addition of one unit matrix; then,
there are the orbits $3-7$ with two or one arbitrary parameter;
there are also orbits $8,\,9$ with two rational parameters, each
orbit have the length depending on the common denominator of its
parameters (we write an estimation of the length instead of the
exact formula): $4\, denominator^2 / \pi^2 < length <
denominator^2 / 2 + 1$;
and there are the orbits $10-131$  with  the following explicit
values. The list is presented in the Table 1.

\begin{center}Table 1: List of 4-signatures which generate finite orbits\end{center}
\[\ba{|c|c||c|c|c|c||c|c|c|c|} \hline
\sharp & length &\theta_1 &\theta_2 &\theta_3 &\theta_4 &\sigma_{12} &\sigma_{23} &\sigma_{13} &\sigma_{24}\\
\hline 1 & &x & y & z &x\!+\!y\!+\!z& x+y & y+z & x+z & x+z\\
\hline 2 & 1 &x & y & z & 0 & z & x & y & y\\
\hline 3 & 2  &x & 1/2 & y & 1/2 & 1/2 & 1/2 & x+y & x-y\\ \hline
4 & 2  &x & y & x & y+1 & 1/2 & 1/2 & 2 x & 2 y+1\\
 \hline
5 & 3 &2 x & x & x & 2/3 & 1/2 & 1/3 & 1/2 & 3 x\\
 \hline
 6 & 4 &x & x & x & 3 x+1 & 1/3 & 1/3 & 1/3 & 4
x+1\\ \hline
7 & 4 &x & x & x & 1/2 & 1/3 & 1/3 & 1/3 & 2 x\\
\hline 8 & & 1/2 & 1/2 & 1/2 & 1/2 & z & y & z\!+\!y\!+\!1 & z\!-\!y\!+\!1\\
\hline 9 & &0 & 0 & 0 & 1 & z & y & z\!+\!y\!+\!1 & z\!-\!y\!+\!1\\
\hline 10&5 &7/15 & 2/5 & 13/15 & 2/5 & 1/2 & 1/2 & 1/2 & 1/2\\
\hline 11&5 &1/5 & 2/5 & 2/5 & 1/5 & 1/3 & 0 & 1/2 & 1/2\\ \hline
12&5 &2/5 & 1/3 & 4/5 & 1/3 & 1/2 & 1/2 & 1/2 & 1/2\\ \hline 13&5
&1 & 1 & 4/5 & 2/5 & 0 & 1/3 & 1/2 & 1/2\\ \hline 14&5 &14/15 &
4/5 & 11/15 & 1/5 & 1/2 & 1/2 & 1/2 & 1/2\\ \hline 15&6 &2/3 & 1/2
& 2/3 & 1/2 & 2/3 & 2/3 & 1/2 & 1/2\\ \hline 16&6 &3/4 & 2/3 & 1/2
& 1/2 & 1/2 & 1/3 & 1/2 & 1/2\\ \hline 17&6 &17/24 & 13/24 & 13/24
& 7/24 & 2/3 & 1/2 & 1/2 & 1/2\\ \hline 18&6 &1 & 5/12 & 1/12 & 0
& 1/4 & 1 & 1/2 &
1/2\\ \hline 19&6 &1/4 & 1/3 & 1/3 & 1/4 & 1/4 & 0 & 1/2 & 1/2\\
\hline 20&6 &19/24 & 19/24 & 23/24 & 1/24 & 1/2 & 1/3 & 1/2 & 1/2\\
\hline 21&6 &1/6 & 1 & 1/6 & 0 & 1/3 & 2/3 & 1/2 & 1/2\\ \hline 22&6 &2/5 & 2/5 & 2/3 & 1/5 & 3/5 & 3/5 & 1/2 & 1/2\\
\hline 23&6 &4/5 & 2/3 & 2/5 & 1/5 & 1/5 & 4/5 & 1/2 & 1/2\\
\hline 24&6
&5/6 & 19/30 & 11/30 & 7/30 & 1/5 & 4/5 & 1/2 & 1/2\\ \hline 25&6 &17/30 & 5/6 & 19/30 & 13/30 & 3/5 & 2/5 & 1/2 & 1/2\\
\hline 26&6 &5/6 & 29/30 & 29/30 & 13/30 & 1/5 &
1/5 & 1/2 & 1/2\\ \hline 27&6 &7/30 & 7/30 & 5/6 & 1/30 & 3/5 & 3/5 & 1/2 & 1/2\\
\hline \ea\]

\[\ba{|c|c||c|c|c|c||c|c|c|c|} \hline
\sharp & length &\theta_1 &\theta_2 &\theta_3 &\theta_4 &\sigma_{12} &\sigma_{23} &\sigma_{13} &\sigma_{24}\\
\hline 28&7 &4/7 & 4/7 & 3/7 & 1/7 & 1/2 & 1/2 & 1/2 & 2/3\\
\hline 29&7 &2/7 & 4/7 & 2/7 & 2/7 & 1/2 & 1/2 & 1/3 & 1/2\\
\hline 30&7 &1/7 & 5/7 & 1/7 & 1/7 & 1/2 & 1/2 & 1/3 & 1/2\\
\hline 31&8 &1/2 & 3/4 & 3/4 & 1/2 & 1/3 & 1/3 & 2/3 &
1/2\\ \hline 32&8 &2/5 & 4/5 & 1/2 & 2/5 & 1/2 & 3/5 & 1/2 & 2/3\\
\hline 33&8 &3/5 & 1/2 & 1/5 & 1/5 & 1/3 & 1/2 & 1/2 & 2/3\\
\hline 34&8 &2/3 & 1/2 & 1/3 & 1/4 & 1/2 & 1/2 & 2/3 & 1/2\\
\hline 35&8 &3/8 & 11/24 & 3/8 & 5/24 & 1/2 & 1/2 & 1/2 & 1/3\\
\hline 36&8 &1/4 & 11/20 & 11/20 & 3/20 & 1/2 & 1/3 & 1/2 & 2/3\\
\hline 37&8 &3/4 & 9/20 & 7/20 & 7/20 & 3/5 & 1/2 & 2/3 & 1/2\\
\hline 38&8 &1/20 & 3/4 & 7/20 & 1/20 & 1/2 & 1/3 & 2/3 & 1/2\\
\hline 39&8 &7/8 & 7/24 & 1/8 & 1/24 & 1/2 & 1/2 & 2/3 & 1/2\\
\hline 40&8 &3/4 & 3/4 & 1 & 0 & 2/3 & 2/3 & 1/2 & 2/3\\ \hline
41&8 &3/4 & 3/20 & 3/20 & 1/20 & 1/2 & 3/5 & 1/3
& 1/2\\
\hline 42&9 &8/15 & 8/15 & 11/15 & 7/15 & 4/5 & 2/5 & 2/5 & 2/5\\
\hline 43&9 &8/15 & 1/15 & 14/15 & 1/15 & 4/5 & 4/5 & 2/5 &
1/3\\ \hline 44&9 &3/5 & 3/5 & 2/3 & 2/5 & 3/5 & 1/5 & 2/5 & 2/5\\
\hline 45&9 &1/3 & 4/5 & 4/5 & 1/5 & 4/5 & 1/5 & 2/3 & 3/5\\
\hline 46&9 &4/15 & 4/15 & 4/15 & 2/15 & 1/5 & 1/5 & 2/5 & 1/3\\
\hline 47&9 &2/15 & 2/15 & 13/15 & 1/15 & 3/5 & 1/5 & 2/5 & 2/5\\
\hline 48&10 &3/5 & 1/2 & 1/2 & 1/5 & 2/3 & 1/2 & 1/2 & 1/2\\
\hline 49&10 &9/10 & 17/30 & 17/30 & 13/30 & 1/2 & 1/2 & 1/2 & 2/5\\
\hline 50&10 &3/5 & 3/5 & 7/10 & 3/10 & 1/2 & 2/3 & 1/2 & 1/2\\
\hline 51&10 &1/3 & 2/5 & 1/3 & 1/3 & 1/2 & 1/2 & 1/5 & 1/2\\
\hline 52&10 &11/30 & 11/30 & 11/30 & 3/10 & 1/2 & 1/2 & 1/2 & 1/5\\
\hline 53&10 &2/3 & 4/5 & 2/3 & 1/3 & 1/2 & 1/2 & 3/5 & 1/2\\
\hline 54&10 &29/30 & 7/10 & 29/30 & 1/30 & 1/2 & 1/2 & 1/2 & 1/5\\
\hline 55&10 &9/10 & 23/30 & 23/30 & 7/30 & 1/2 & 1/2 & 2/5 &
1/2\\ \hline 56&10 &1/5 & 1/5 & 9/10 & 1/10 & 1/2 & 2/3 & 1/2 & 1/2\\
\hline 57&10 &3/5 & 2/5 & 3/5 & 2/5 & 2/5 & 2/5 & 1/3 & 3/5\\
\hline 58&10 &1 & 1 & 1 & 3/5 & 0 & 1/5
& 1/3 & 1/3\\ \hline 59&10 &4/5 & 4/5 & 4/5 & 4/5 & 1/5 & 0 & 1/3 & 1/3\\
\hline 60&10 &1 & 1 & 1 & 1/5 & 3/5 & 3/5 & 3/5 & 1/3\\
\hline \ea\]

 \[\ba{|c|c||c|c|c|c||c|c|c|c|} \hline
\sharp & length &\theta_1 &\theta_2 &\theta_3 &\theta_4 &\sigma_{12} &\sigma_{23} &\sigma_{13} &\sigma_{24}\\
\hline 61&12 &1/2 & 1/2 & 1/2 & 1/3 & 1/2 & 3/4 & 2/3 & 1/3\\
\hline 62&12 &1/2 & 2/3 & 2/3 & 1/2 & 1/4 & 1/2 & 1/2 & 1/2\\
\hline 63&12 &1/2 & 2/3 & 3/5 & 2/5 & 2/5 & 2/3 & 1/2 & 1/2\\
\hline 64&12 &3/5 & 1/2 & 1/3 & 1/5 & 1/2 & 1/2 & 3/5 & 1/2\\
\hline 65&12 &1/2 & 4/5 & 4/5 & 1/3 & 2/3 & 1/5 & 1/2 & 1/2\\
\hline 66&12 &7/12 & 7/12 & 29/60 & 19/60 & 3/5 & 1/3 & 1/2 & 1/2\\
\hline 67&12 &37/60 & 31/60 & 41/60 & 13/60 & 1/2 & 1/2 & 1/2 & 2/5\\
\hline 68&12 &7/10 & 17/30 & 3/10 & 7/30 & 2/5 & 3/5 & 4/5 & 1/2\\
\hline 69&12 &5/12 & 7/12 & 5/12 & 5/12 & 1/2 & 3/4 & 2/3 & 1/3\\
\hline 70&12 &7/12 & 53/60 & 43/60 & 7/12 & 1/3 & 1/5 & 1/2 & 1/2\\
\hline 71&12 &1/3 & 4/5 & 3/5 & 1/3 & 3/5 & 1/2 & 3/5 & 4/5\\
\hline 72&12 &11/12 & 11/12 & 37/60 & 13/60 & 1/5 & 2/3 & 1/2 & 1/2\\
\hline 73&12 &29/30 & 9/10 & 9/10 & 11/30 & 2/5 & 1/2 & 2/5 & 1/5\\
\hline 74&12 &17/60 & 11/60 & 59/60 & 7/60 & 1/2 & 1/2 & 2/5 & 1/2\\
\hline 75&12 &1/12 & 11/12 & 11/60 & 1/60 & 2/5 & 2/3 &
1/2 & 1/2\\ \hline 76&12 &1 & 1 & 5/6 & 1/6 & 1/2 & 3/4 & 1/2 & 1/2\\
\hline 77&12 &11/12 & 11/12 & 11/12 & 1/12 & 1/4 & 1/2 & 1/3 & 2/3\\
\hline 78&15 &8/15 & 7/15 & 4/5 & 7/15 & 2/5 & 2/5 & 1/2 & 1/2\\
\hline 79&15 &13/15 & 1 & 1 & 7/15 & 1/5 & 0 & 1/2 & 1/2\\ \hline
80&15 &1/3 & 3/5 & 3/5 & 1/3 & 3/5 & 0 & 1/2 & 1/2\\ \hline 81&15
&1/3 & 1/3 & 3/5 & 1/3 & 3/5 & 3/5 & 1/2 & 1/2\\ \hline 82&15
&4/15 & 4/15 & 2/5 & 4/15 & 1/5 & 1/5 & 1/2 & 1/2\\ \hline 83&15
&14/15 & 14/15 & 3/5 & 1/15 & 1/5 & 4/5 & 1/2 & 1/2\\ \hline 84&15
&4/5 & 2/3 & 1/3 & 1/3 & 1/5 & 4/5 & 1/2 & 1/2\\ \hline 85&15 &4/5
& 2/3 & 1/3 & 1/5 & 1/5 & 1 & 1/2 & 1/2\\ \hline 86&15 &14/15 &
4/15 & 0 & 0 & 0 & 3/5 & 1/2 & 1/2\\ \hline 87&15 &13/15 & 1/5 &
2/15 & 2/15 & 3/5 & 2/5 & 1/2
& 1/2\\ \hline 88&16 &3/4 & 1/2 & 1/2 & 1/2 & 2/3 & 1/3 & 1/2 & 1/3\\
\hline 89&16 &5/8 & 5/8 & 5/8 & 3/8 & 1/3 & 2/3 & 1/3 & 1/2\\
\hline 90&16 &7/8 & 7/8 & 7/8 & 1/8 & 1/3 & 2/3 & 1/3 & 1/2\\
\hline 91&18 &29/42 & 23/42 & 23/42 & 19/42 & 1/2 & 1/3 & 2/3 & 2/7\\
\hline 92&18 &31/42 & 31/42 & 23/42 & 11/42 & 1/2 & 4/7 & 4/7 & 2/3\\
\hline 93&18 &4/7 & 3/7 & 3/7 & 1/3 & 2/3 & 1/3 & 2/7 & 1/2\\
\hline 94&18 &17/42 & 17/42 & 17/42 & 5/42 & 1/2 & 3/7 & 3/7 & 1/3\\
\hline 95&18 &1/42 & 41/42 & 17/42 & 1/42 & 6/7 & 1/7 & 1/2 & 3/7\\
\hline 96&18 &2/3 & 2/3 & 2/3 & 2/3 & 0 & 1/5 & 3/5 & 3/5\\ \hline
97&18 &2/7 & 2/7 & 1/3 & 2/7 & 1/7 & 1/7 & 1/2 & 4/7\\ \hline
98&18 &6/7 & 6/7 & 6/7 & 1/3 & 1/2 & 3/7 & 3/7 & 1/3\\ \hline
99&18 &1 & 1/3 & 0 & 0 & 1/5 & 1 & 2/5 & 2/5\\ \hline 100&18
&29/42 & 29/42 & 13/42 & 11/42 & 1/7 & 6/7 & 1/2 & 3/7\\ \hline
101&18 &37/42 & 37/42 & 37/42 & 1/42 & 1/2 & 1/3 & 1/3 & 5/7\\
\hline \ea\]

 \[\ba{|c|c||c|c|c|c||c|c|c|c|} \hline
\sharp & length &\theta_1 &\theta_2 &\theta_3 &\theta_4 &\sigma_{12} &\sigma_{23} &\sigma_{13} &\sigma_{24}\\
\hline 102&20 &3/5 & 1/2 & 1/2 & 2/5 & 4/5 & 2/5 & 1/2 & 1/3\\
\hline 103&20 &1/2 & 1/2 & 4/5 & 1/5 & 2/3 & 3/5 & 1/2 & 3/5\\
\hline 104&20 &3/5 & 1/2 & 1/3 & 1/3 & 1/2 & 1/3 & 1/2 & 3/5\\
\hline 105&20 &1/2 & 1/3 & 4/5 & 1/3 & 1/2 & 3/5 & 1/3 & 1/2\\
\hline 106&20 &13/20 & 13/20 & 29/60 & 11/60 & 1/2 & 1/2 & 2/3 & 3/5\\
\hline 107&20 &43/60 & 37/60 & 11/20 & 11/20 & 1/2 & 2/3 & 1/2 & 2/5\\
\hline 108&20 &17/20 & 19/60 & 3/20 & 1/60 & 1/2 & 3/5 & 1/2 & 2/3\\
\hline 109&20 &1 & 1 & 7/10 & 3/10 & 1/3 & 3/5 & 1/2 & 2/5\\
\hline 110&20 &1/20 & 47/60 & 7/60 & 1/20 & 2/3 & 1/2 & 2/5 & 1/2\\
\hline 111&20 &1 & 9/10 & 9/10 & 0 & 1/5 & 2/5 & 2/3 & 1/2\\
\hline 112&24 &1/2 & 1/3 & 2/3 & 1/3 & 1/2 & 2/5 & 1/2 & 1/3\\
\hline 113&24 &7/12 & 5/12 & 7/12 & 1/4 & 2/5 & 1/2 & 1/3 & 1/2\\
\hline 114&24 &3/4 & 1/12 & 1/12 & 1/12 & 1/2 & 3/5 & 1/3 & 1/2\\
\hline 115&30 &1/2 & 1/2 & 2/5 & 1/3 & 1/2 & 1/5 & 1/2 &
1/2\\ \hline 116&30 &1/2 & 1/2 & 2/3 & 1/5 & 1/2 & 2/5 & 1/2 & 1/2\\
\hline 117&30 &8/15 & 19/30 & 19/30 & 7/15 & 1/5 & 1/2 & 1/2 & 1/2\\
\hline 118&30 &11/15 & 11/15 & 17/30 & 13/30 & 1/2 & 3/5 & 1/2 & 1/2\\
\hline 119&30 &1/15 & 23/30 & 7/30 & 1/15 & 2/5 & 1/2 & 1/2 & 1/2\\
\hline 120&30 &13/15 & 2/15 & 1/30 & 1/30 & 1/2 & 4/5 & 1/2 & 1/2\\
\hline 121&36 &1/2 & 2/3 & 2/3 & 1/2 & 1/3 & 3/5 & 1/2 & 2/5\\
\hline 122&36 &1 & 1 & 5/6 & 1/6 & 3/5 & 1/3 & 1/2 & 2/5\\
\hline 123&40 &1/2 & 1/2 & 1/2 & 2/5 & 1/2 & 4/5 & 2/3 & 1/3\\
\hline 124&40 &1/2 & 4/5 & 1/2 & 1/2 & 2/5 & 2/3 & 2/5 & 1/2\\
\hline 125&40 &11/20 & 9/20 & 9/20 & 9/20 & 1/2 & 1/5 & 1/3 & 2/3\\
\hline 126&40 &13/20 & 13/20 & 13/20 & 7/20 & 2/3 & 3/5 & 3/5 & 1/2\\
\hline 127&40 &17/20 & 17/20 & 17/20 & 3/20 & 1/2 & 1/3 & 3/5 & 2/5\\
\hline 128&40 &19/20 & 19/20 & 19/20 & 1/20 & 1/2 & 4/5 & 1/3 & 2/3\\
\hline 129&72 &1/2 & 1/2 & 1/2 & 1/3 & 4/5 & 1/2 & 2/5 & 3/5\\
\hline 130&72 &5/12 & 7/12 & 5/12 & 5/12 & 1/2 & 4/5 & 3/5 & 2/5\\
\hline 131&72 &1/12 & 1/12 & 11/12 & 1/12 & 1/2 & 1/5 & 2/5 & 3/5\\
\hline \ea\]

\section{Construction of ${\mathcal M}^{(n)}$'s of higher order}

{\lemma{$\,$}\label{reduction_finite}}
 For each ${\mathcal M_F}^{(n)}$
which will be notated as
\[M_1,\, M_2,\, M_3\,...\, M_n\]
its reduction
\[(M_1\cdot M_2),\, M_3\,...\, M_n\]
is a ${\mathcal M_F}^{(n-1)}$ --- it also belongs to a finite orbit.

\textbf{Proof:} Here we use the long form of the ${\mathcal M}^{(n)}$.
The starting point of the orbit we denote

\[M_1,\, M_2,\,...\, M_n,\quad N_1 = 1,\,N_2 = 2,\,...\, N_n = n.\]

Having all elements of the orbit, we select a subset with the
following condition: for such $k$ that $N_k = 1$ we require that
$N_{k+1} = 2$ (observe that the index $k$ is defined modulo $n$).

For the remaining tuples ${\mathcal M}^{(n)}$'s we allow only the
following braid group actions: any ${\mathcal B}_{m,m+1}$ and
${\mathcal B}_{m+1,m}$ for $m \neq k-1,\,k,\,k+1$;  compositions
${\mathcal B}_{k,k+1}{\mathcal B}_{k-1,k}$, ${\mathcal
B}_{k+1,k}{\mathcal B}_{k+2,k+1}$, ${\mathcal B}_{k,k+1}{\mathcal
B}_{k+1,k+2}$ and ${\mathcal B}_{k+1,k}{\mathcal B}_{k,k-1}$,
where $k$ is such value that $N_k = 1$.

These allowed braid group actions preserve the condition from the
previous paragraph: $N_k = 1 \rightarrow N_{k+1} = 2$.

Next, we can do the reduction of elements of this set, joining the
matrices $M_k,\, M_{k+1}$ for which $N_k = 1,\,N_{k+1} = 2$ into
their product $M_k\cdot M_{k+1}$.

To describe such reduction more simply we use the fact that the
set is closed under cyclical permutation (see \ref{symgroup1}),
and define the reduction only for such ${\mathcal M}^{(n)}$'s that
$N_{n-1} = 1$, $N_n = 2$. The reduction operation will be notated
as $r_{k,k+1}$,

\[r_{k,k+1}\{...M_k,\,M_{k+1},...\} \rightarrow \{...M_k\cdot M_{k+1},...\},\]
\[r_{n,1}\{M_1,...M_n\} \rightarrow \{M_n\cdot M_1,...\},\]
so we have:
\[\ba{l}
{\mathcal B}_{m,m+1}\, r_{n-1,n}{\mathcal M}^{(n)} = r_{n-1,n}
{\mathcal
B}_{m,m+1}{\mathcal M}^{(n)}, \quad m < n-2,\\
{\mathcal B}_{m+1,m}\, r_{n-1,n}{\mathcal M}^{(n)} = r_{n-1,n}
{\mathcal B}_{m+1,m}{\mathcal M}^{(n)},\quad m <
n-2,\\
{\mathcal B}_{n-2,n-1}\,r_{n-1,n}{\mathcal M}^{(n)} =
r_{n-2,n-1}{\mathcal B}_{n-1,n}{\mathcal
B}_{n-2,n-1}{\mathcal M}^{(n)},\\
{\mathcal B}_{n-1,n-2}\,r_{n-2, n-1}{\mathcal M}^{(n)} =
r_{n-1,n}{\mathcal B}_{n-1,n-2}{\mathcal
B}_{n,n-1}{\mathcal M}^{(n)},\\
{\mathcal B}_{n-1,1}\, r_{n, 1}{\mathcal M}^{(n)}= r_{n-1, n}
{\mathcal B}_{n,1}{\mathcal
B}_{n-1,n}{\mathcal M}^{(n)},\\
{\mathcal B}_{1,n-1}\, r_{n-1,n}{\mathcal M}^{(n)} = r_{n,1}
{\mathcal B}_{n,n-1}{\mathcal B}_{1,n}{\mathcal M}^{(n)}, \ea\]
and  for each obtained ${\mathcal M}^{(n-1)}$ add all its copies
obtained by cyclical permutation.

Now we have a set of the tuples ${\mathcal M}^{(n-1)}$'s, finite and closed
under all braid group actions.

It is for sure either a finite orbit or a set of several finite
orbits.

Lemma is proven. $\square$

{\lemma{$\,$}\label{induction_mo}} As a corollary of the previous
lemma, every ${\mathcal M_F}^{(n)}$ can be constructed from two
${\mathcal M_F}^{(n-1)}$'s:

\textbf{Proof:} We take any two tuples ${\mathcal M_F}^{(n-1)}$, which
coincide with all matrices, except for two neighboring ones. In one
${\mathcal M_F}^{(n-1)}$, we call these two matrices $A$ and $B$
\[A,\, B,\, M_3\,...\,M_{n-1}\]
and in another tuple we call them $C$ and $C^{-1}A\,B$ (so that
the
product of both matrices is equal to the one in  the first tuple)
\[C,\,C^{-1}A B,\, M_3\,...\, M_{n-1}.\]
We construct from them the tuple ${\mathcal M}^{(n)}$:
\[C,\,C^{-1}A,\, B,\, M_3\,...\, M_{n-1}.\]
However, we are not yet sure that it belongs to a finite orbit; nevertheless
every tuple  ${\mathcal M_F}^{(n)}$ can be constructed in such way. $\square$

Using all pairs of ${\mathcal M_F}^{(n-1)}$'s we can get a list of
the tuples ${\mathcal M}^{(n)}$'s which is a complete list of the
candidates for ${\mathcal M_F}^{(n)}$'s. Then we check which of
them really generate finite orbits. It gives us a hope to get the
complete list of ${\mathcal M_F}^{(n)}$'s by a finite procedure.

For $n = 1,\,2,\,3$ the problem of classification of finite orbits
is trivial: every ${\mathcal M}^{(1)}$, ${\mathcal M}^{(2)}$ or ${\mathcal
M}^{(3)}$ generates a finite orbit of length $1$.

For $n = 4$ this problem was solved in our paper about Painleve-VI
equation \cite{painleve_6}.

Therefore, using the list of finite orbits of ${\mathcal M}^{(4)}$'s, we can
obtain the list of finite orbits of ${\mathcal M}^{(5)}$'s, and then
using the list of ${\mathcal M}^{(5)}$'s, obtain the corresponding list for ${\mathcal
M}^{(6)}$'s et cetera.

In this paper we developed the algorithm for exact and exhaustive
search of ${\mathcal M}^{(n)}$ tuples. Our aim was to make this
algorithm in such a way that it needs only simple arithmetic and
algebra. This allowed us designing a special computer program to
perform this search for $n=5$, because it would be too many
calculation for a human.

\section{Example of construction of ${\mathcal M_F}^{(5)}$ with direct using of
matrices}

Let us assume that in the list below the following five ${\mathcal
M_F}^{(4)}$'s occur. Let us call them $A,B,C,D,E$:

\[A: \quad \left[A_1,\quad A_2,\quad A_3,\quad A_4\right] = \]
\[= \left[\left(\ba{cc}1&0\\0&1\ea\right),\quad\left(\ba{cc}1&1\\-1&0\ea\right),\quad
\left(\ba{cc}0&-i\\-i&1\ea\right),\quad\left(\ba{cc}i&i-1\\0&-i\ea\right)\right];\]

\[B: \quad \left[B_1,\quad B_2,\quad B_3,\quad B_4\right] = \]
\[= \left[\left(\ba{cc}1&1\\-1&0\ea\right),\quad\left(\ba{cc}1&0\\0&1\ea\right),\quad
\left(\ba{cc}0&-i\\-i&1\ea\right),\quad\left(\ba{cc}i&i-1\\0&-i\ea\right)\right];\]

\[C: \quad \left[C_1,\quad C_2,\quad C_3,\quad C_4\right] = \]
\[= \left[\left(\ba{cc}1&1\\-1&0\ea\right),\quad\left(\ba{cc}0&-1\\1&1\ea\right),\quad
\left(\ba{cc}-i&1-i\\0&i\ea\right),\quad\left(\ba{cc}i&i-1\\0&-i\ea\right)\right];\]

\[D: \quad \left[D_1,\quad D_2,\quad D_3,\quad D_4\right] = \]
\[= \left[\left(\ba{cc}1&1\\-1&0\ea\right),\quad\left(\ba{cc}0&-1\\1&1\ea\right),\quad\left(\ba{cc}1&1\\-1&0\ea\right),\quad
\left(\ba{cc}0&-1\\1&1\ea\right)\right];\]

\[E: \quad \left[E_1,\quad E_2,\quad E_3,\quad E_4\right] = \]
\[= \left[\left(\ba{cc}0&-1\\1&1\ea\right),\quad\left(\ba{cc}1&1\\-1&0\ea\right),\quad
\left(\ba{cc}0&-i\\-i&1\ea\right),\quad\left(\ba{cc}1&i\\i&0\ea\right)\right].\]

In fact, $A$ and $B$ belong to one orbit of  the length $1$ (orbit
number $2$ in Table 1), $C$ belongs to an orbit of the length $6$
(a symmetry of the orbit number $15$), $D$ belongs to another
orbit of the length $1$ (orbit number $1$, all matrices in the
tuple $D$ can be diagonalized simultaneously), and $E$ -- to the
orbit of length $3$ (a symmetry of the orbit number $5$).

We will obtain  ${\mathcal M}^{(5)}$ from these tuples
by the procedure given below.

Due to the construction principle exposed in the previous chapter, we
 notice that $A_3 = B_3$ and $A_4 = B_4$, so the tuples
$A$ and $B$ differ only by two first matrices.

Then we can perform {\it induction} of the tuple $A$ into
tuple $F$, meaning that $A_1$ is $F_1\cdot F_2$, renaming $A_2$ to
$F_3$, $A_3$ to $F_4$ and $A_4$ to $F_5$:
\[[A_1,\quad A_2, \quad A_3, \quad A_4] = [F_1\cdot F_2,\quad F_3,\quad F_4,\quad F_5].\]
But it doesn't give us information about $F_1$ and $F_2$. To
obtain it we can use information from the tuple $B$:

\[[B_1,\quad B_2, \quad B_3, \quad B_4] = [F_1,\quad F_2 \cdot F_3,\quad F_4,\quad F_5].\]

In such a way we have found all the matrices $F$:

\be\label{example}\left[\left(\ba{cc}1&1\\-1&0\ea\right),\quad\left(\ba{cc}0&-1\\1&1\ea\right),\quad\left(\ba{cc}1&1\\-1&0\ea\right),\quad
\left(\ba{cc}0&-i\\-i&1\ea\right),\quad\left(\ba{cc}\,\,i\,\,&i-1\\0&-i\ea\right)\right].\ee

But we can also use   $C$, $D$ and $E$ for the next {\it
inductions}:

\[[C_1,\quad C_2, \quad C_3, \quad C_4] = [F_1,\quad F_2,\quad F_3 \cdot F_4,\quad F_5],\]
\[[D_1,\quad D_2, \quad D_3, \quad D_4] = [F_1,\quad F_2,\quad F_3,\quad F_4 \cdot F_5],\]
\[[E_1,\quad E_2, \quad E_3, \quad E_4] = [F_2,\quad F_3,\quad F_4,\quad F_5\cdot F_1].\]
Therefore despite  using the pair $A$ and $B$ we could use
similarly any of the pairs $B$ and $C$, $C$ and $D$, $D$ and $E$
or $E$ and $A$.

 Observe that for the signature
formalism which will be described in the next chapter, we will
need all the above written five 4-tuples of matrices.

The previous formulae still do not prove that $F$ generates a finite
orbit. But due to the Lemma \ref{induction_mo} we are sure that
every ${\mathcal M_F}^{(5)}$ can be constructed in a similar way.

For the $F$ case we can check explicitly whether the generated
orbit is finite or infinite. It turns out to be a finite orbit of
length 16 (orbit number $9$ in Table 9), further it will be called
a tetrahedral type orbit.

In order to obtain an exhaustive list of the tuples ${\mathcal M_F}^{(5)}$ we
must repeat the procedure described above for each five ${\mathcal
M_F}^{(4)}$ from the list of all possible ${\mathcal M_F}^{(4)}$'s.

The method of classification of finite ${\mathcal M}^{(5)}$ orbits,
described in this example, cannot be used in practice, because the
set of ${\mathcal M_F}^{(4)}$ is infinite.

The set of ${\mathcal M_F}^{(4)}$ can be described as a finite list
using free parameters. Now we will repeat the procedure of this
example, using  ${\mathcal M_F}^{(4)}$'s in a
parametric form as a source of our construction.

First, assume that written below tuple ${\mathcal M_F}^{(4)}$ with three
parameters occurs in the list. We will call it $A$, and we will use it as the first
stage of construction of $F$ tuple:

\[A: \quad \left[A_1,\quad A_2,\quad A_3,\quad A_4\right] = \]
\[= \quad \left[F_1\cdot F_2,\quad F_3,\quad F_4,\quad F_5\right] = \]
\[= \left[\left(\ba{cc}1&0\\0&1\ea\right),\quad\left(\ba{cc} 2\cos(\pi x) &1\\-1&0\ea\right),\quad
\left(\ba{cc}0&-\exp(i\pi z)\\\exp(-i\pi z) & 2\cos(\pi
y)\ea\right),\right.\]
\[\left.\left(\ba{cc}\exp(i\pi z)\quad &2\exp(i\pi
z)\cos(\pi x)-2\cos(\pi y)\\0&\exp(-i\pi z))\ea\right)\right].\]

Next, we find in the list another tuple, also  three-parameteric:

\[\left[\left(\ba{cc} 2\cos(\pi x) &1\\-1&0\ea\right),\quad
\left(\ba{cc}1&0\\0&1\ea\right),\quad\left(\ba{cc}0&-\exp(i\pi
z)\\\exp(-i\pi z) & 2\cos(\pi y)\ea\right),\right.\]
\[\left.\left(\ba{cc}\exp(i\pi z)\quad &2\exp(i\pi
z)\cos(\pi x)-2\cos(\pi y)\\0&\exp(-i\pi z))\ea\right)\right].\]

We are going to call it $B$, but we must rename the parameters in it,
to avoid collision of  our notations:

\[B: \quad \left[B_1,\quad B_2,\quad B_3,\quad B_4\right] = \]
\[=  \left[\left(\ba{cc} 2\cos(\pi a) &1\\-1&0\ea\right),\quad
\left(\ba{cc}1&0\\0&1\ea\right),\quad\left(\ba{cc}0&-\exp(i\pi
c)\\\exp(-i\pi c) & 2\cos(\pi b)\ea\right),\right.\]
\[\left.\left(\ba{cc}\exp(i\pi c)\quad &2\exp(i\pi
c)\cos(\pi a)-2\cos(\pi b)\\0&\exp(-i\pi c))\ea\right)\right].\]

Now we have to find the condition on the parameters that lead to the equalities $B_3 = F_4$
and $B_4 = F_5$ up to common conjugation.

It gives us a discrete set of possibilities:

\[a = \pm x,\quad b = \pm y,\quad c = \pm z,\]

and we will choose one of them:

\[a = -x,\quad b = -y,\quad c = -z.\]

Therefore the  tuple $B$ turns into the following expression:

\[B: \quad \left[B_1,\quad B_2,\quad B_3,\quad B_4\right] = \]
\[=  \left[\left(\ba{cc} 2\cos(\pi x) &1\\-1&0\ea\right),\quad
\left(\ba{cc}1&0\\0&1\ea\right),\quad\left(\ba{cc}0&-\exp(-i\pi
z)\\\exp(i\pi z) & 2\cos(\pi y)\ea\right),\right.\]
\[\left.\left(\ba{cc}\exp(-i\pi z)\quad &2\exp(-i\pi
z)\cos(\pi x)-2\cos(\pi y)\\0&\exp(i\pi z))\ea\right)\right].\]

Now in order to make $B_3$ to coincide with $F_4$ and $B_4$
with $F_5$ we will do a common conjugation
\[B_\nu \rightarrow \left(\ba{cc}\exp(i\pi z)\cos(\pi x) - \cos(\pi y)&i\sin(\pi z)\\-i\sin(\pi z) & \exp(-i\pi z)\cos(\pi x) - \cos(\pi
y)\ea\right)\cdot B_\nu\cdot\] \[\cdot \left(\ba{cc}\exp(i\pi
z)\cos(\pi x) - \cos(\pi y)&i\sin(\pi z)\\-i\sin(\pi z) &
\exp(-i\pi z)\cos(\pi x) - \cos(\pi y)\ea\right)^{-1}, \]
and obtain the following expression:
\[B: \quad \left[B_1,\quad B_2,\quad B_3,\quad B_4\right] = \]
\[=  \left[\left(\ba{cc} 2\cos(\pi x) &1\\-1&0\ea\right),\quad
\left(\ba{cc}1&0\\0&1\ea\right),\quad\left(\ba{cc}0&-\exp(i\pi
z)\\\exp(-i\pi z) & 2\cos(\pi y)\ea\right),\right.\]
\[\left.\left(\ba{cc}\exp(i\pi z)\quad &2\exp(i\pi
z)\cos(\pi x)-2\cos(\pi y)\\0&\exp(-i\pi z))\ea\right)\right].\]

Finally, we can put $F_1 = B_1$ and $F_2\cdot F_3 = B_2$, that is why
$F_2 = B_2 \cdot F_3^{-1}$, and  we obtain
\[F: \quad \left[F_1,\quad F_2,\quad F_3,\quad F_4,\quad F_5\right] = \]
\[=  \left[\left(\ba{cc} 2\cos(\pi x) &1\\-1&0\ea\right),\quad \left(\ba{cc}
0 &-1\\1&2\cos(\pi x)\ea\right),\right.\]\[\left(\ba{cc} 2\cos(\pi
x) &1\\-1&0\ea\right),\quad \left(\ba{cc}0&-\exp(i\pi
z)\\\exp(-i\pi z) & 2\cos(\pi y)\ea\right),\]
\[\left.\left(\ba{cc}\exp(i\pi z)\quad &2\exp(i\pi
z)\cos(\pi x)-2\cos(\pi y)\\0&\exp(-i\pi z))\ea\right)\right].\]

On the next step, we will get from the list one more tuple ${\mathcal M_F}^{(4)}$
and call it $C$:
\[C: \quad \left[C_1,\quad C_2,\quad C_3,\quad C_4\right] = \]
\[= \left[\left(\ba{cc}1&1\\-1&0\ea\right),\quad\left(\ba{cc}0&-1\\1&1\ea\right),\quad
\left(\ba{cc}-i&1-i\\0&i\ea\right),\quad\left(\ba{cc}i&i-1\\0&-i\ea\right)\right],\]
and will require the conditions $C_1 = F_1$, $C_2 = F_2$, $C_3 =
F_3\cdot F_4$, $C_4 = F_5$.
 To achieve  this we don't need any common conjugation and must fix only all the
parameters:
\[x = 1/3,\quad y = 1/3,\quad z = 1/2.\]

Therefore in this way we obtain:

\[F: \quad \left[F_1,\quad F_2,\quad F_3,\quad F_4,\quad F_5\right] = \]
\[=\left[\left(\ba{cc}1&1\\-1&0\ea\right),\quad\left(\ba{cc}0&-1\\1&1\ea\right),\quad\left(\ba{cc}1&1\\-1&0\ea\right),\quad
\left(\ba{cc}0&-i\\-i&1\ea\right),\quad\left(\ba{cc}i&i-1\\0&-i\ea\right)\right].\]

Further on, we take one more 4-tuple and call it $D$:

\[D: \quad \left[D_1,\quad D_2,\quad D_3,\quad D_4\right] = \]
\[= \left[\left(\ba{cc}\exp(i\pi f)&0\\k&\exp(-i\pi f)\ea\right),
\quad\left(\ba{cc}\exp(i\pi g)&0\\l&\exp(-i\pi
g)\ea\right),\right.\]\[\left.\left(\ba{cc}\exp(i\pi
h)&0\\m&\exp(-i\pi h)\ea\right),\quad \left(\ba{cc}\exp(-i f-i g-i
h)&0\\n&\exp(i f+i g + i h)\ea\right)\right].\]

Now we have to provide the equalities $D_1 = F_1$, $D_2 = F_2$, $D_3 = F_3$.

From the equality $Tr\, D_1 = Tr\, F_1$ up to a common conjugation
we obtain $f = 1/3$. Further, from the equalities $D_1\cdot D_2 =
F_1\cdot F_2 = 1$ and $D_2\cdot D_3 = F_2\cdot F_3 = 1$ we get $g
= -1/3$, $h = 1/3$, $l = -k$, $m = k$, $n = -k$. Therefore in this
way we obtain

\[D: \quad \left[D_1,\quad D_2,\quad D_3,\quad D_4\right] = \]
\[= \left[\left(\ba{cc}\frac{1+i\sqrt{3}}{2}&0\\k&\frac{1-i\sqrt{3}}{2}\ea\right),
\quad\left(\ba{cc}\frac{1-i\sqrt{3}}{2}&0\\-k&\frac{1+i\sqrt{3}}{2}\ea\right),\right.\]
\[\left.\left(\ba{cc}\frac{1+i\sqrt{3}}{2}&0\\k&\frac{1-i\sqrt{3}}{2}\ea\right),\quad
\left(\ba{cc}\frac{1-i\sqrt{3}}{2}&0\\-k&\frac{1+i\sqrt{3}}{2}\ea\right)\right].\]

Now in order to make  $D_1$ and $F_1$  equal  we perform a common
conjugation:
\[D_\nu\rightarrow\left(\ba{cc}0 & 1 \\ k & -\frac{1+i
\sqrt{3}}{2}\ea\right)\cdot D_\nu\cdot \left(\ba{cc}0 & 1 \\ k &
-\frac{1+i \sqrt{3}}{2}\ea\right)^{-1}\]
and obtain
\[D: \quad \left[D_1,\quad D_2,\quad D_3,\quad D_4\right] = \]
\[= \left[\left(\ba{cc}1&1\\-1&0\ea\right),
\quad\left(\ba{cc}0&-1\\1&1\ea\right),\quad\left(\ba{cc}1&1\\-1&0\ea\right),\quad
\left(\ba{cc}0&-1\\1&1\ea\right)\right].\]

In such a way the conditions $D_1 = F_1$, $D_2 = F_2$, $D_3 = F_3$, $D_4 =
F_4\cdot F_5$ are satisfied.

Finally we can find 4-tuple the written below call it $E$:

\[E: \quad \left[E_1,\quad E_2,\quad E_3,\quad E_4\right] = \]
\[=\left[\left(\ba{cc}-U^2&-U-U^{-1}\\0&-U^{-2}\ea\right),\quad
\left(\ba{cc}U&1\\0&U^{-1}\ea\right),\quad
\left(\ba{cc}0&U\\-U^{-1}&1\ea\right),\right.\]
\[\left.
\left(\ba{cc}-U^{-3}&U^4+U^2+1+U^{-2}\\-U^{-4}&U+U^{-1}+U^{-3}\ea\right)\right],\]
where parameter $U$ can also be defined as $\exp(i\pi u)$.

In order to satisfy the condition $Tr\, E_2 = Tr\, F_2$ we put $U = \exp(i\pi/3)$
and obtain

\[E: \quad \left[E_1,\quad E_2,\quad E_3,\quad E_4\right] = \]
\[=\left[\left(\ba{cc}\frac{1-i\sqrt{3}}{2}&-1\\0&\frac{1+i\sqrt3}{2}\ea\right),\quad
\left(\ba{cc}\frac{1+i\sqrt3}{2}&1\\0&\frac{1-i\sqrt3}{2}\ea\right),\quad
\left(\ba{cc}0&\frac{1+i\sqrt3}{2}\\\frac{-1+i\sqrt3}{2}&1\ea\right),\right.\]
\[\left.
\left(\ba{cc}1&\frac{-1-i\sqrt3}{2}\\\frac{1-i\sqrt3}{2}&0\ea\right)\right].\]

Now we  do a common conjugation

\[E_\nu \rightarrow \left(\ba{cc}  1+i\sqrt3 & -i + \sqrt3 \\ -2 &
(1-i)(1-\sqrt3)\ea\right)\cdot E_\nu\cdot \left(\ba{cc}  1+i\sqrt3
& -i + \sqrt3 \\ -2 & (1-i)(1-\sqrt3)\ea\right)^{-1} \]
and  obtain the equality
\[E: \quad \left[E_1,\quad E_2,\quad E_3,\quad E_4\right] = \]
\[= \left[\left(\ba{cc}0&-1\\1&1\ea\right),\quad\left(\ba{cc}1&1\\-1&0\ea\right),\quad
\left(\ba{cc}0&-i\\-i&1\ea\right),\quad\left(\ba{cc}1&i\\i&0\ea\right)\right].\]
Therefore $E_2 = F_2$, $E_3 = F_3$, $E_4 = F_4$ and $E_1 = F_5\cdot F_1$.

By a procedure described above the full list of ${\mathcal
M_F}^{(5)}$ can be obtained with a finite number of steps, but it
needs too much work to be made manually, and also requires rather
complicated mathematics programming a computer.

In the next chapter we will describe how this work could be
simplified using the signature formalism.

\section{Example of construction of ${\mathcal M_F}^{(5)}$ with signature formalism}

In this chapter we repeat the result of the previous chapter using
the signature formalism.

To begin with, we perform  it with explicit values only, without free
parameters.

Assume that in the list of ${\mathcal M_F}^{(4)}$'s signatures the
following five signatures, which will be called  $A,B,C,D,E$,
occur: see Table 2.

\pagebreak

\begin{center}Table 2: 4-signatures for constructing of a 5-signature\end{center}
\[\ba{|c||c|c|c|c|c|}\hline &A&B&C&D&E \\\hline\hline
\theta_1 & 0 &1/3 &1/3 &1/3 &1/3 \\\hline \theta_2 &1/3 &0 &1/3
&1/3 &1/3
\\\hline \theta_3 &1/3 &1/3 &1/2 &1/3 &1/3
\\\hline \theta_4 &1/2 &1/2 &1/2 &1/3 &1/3 \\\hline \sigma_{12} &1/3 &1/3 &0 &0
&0
\\\hline \sigma_{23} &1/2 &1/3 &1/3 &0 &1/2 \\\hline \sigma_{13} &1/3 &1/2 &2/3 &2/3
&1/3
\\\hline \sigma_{24} &1/3 &1/2 &2/3 &2/3 &1/3 \\\hline \ea\]

Now we must perform an induction in each of these signatures, transforming
the 4-signature $A$ into 5-signature $A'$ and so on:
\[[A_1,\quad A_2, \quad A_3, \quad A_4] \rightarrow [A'_1\cdot A'_2,\quad A'_3,\quad A'_4,\quad A'_5],\]
\[[B_1,\quad B_2, \quad B_3, \quad B_4] \rightarrow [B'_1,\quad B'_2 \cdot B'_3,\quad B'_4,\quad B'_5],\]
\[[C_1,\quad C_2, \quad C_3, \quad C_4] \rightarrow [C'_1,\quad C'_2,\quad C'_3 \cdot C'_4,\quad C'_5],\]
\[[D_1,\quad D_2, \quad D_3, \quad D_4] \rightarrow [D'_1,\quad D'_2,\quad D'_3,\quad D'_4 \cdot D'_5],\]
\[[E_1,\quad E_2, \quad E_3, \quad E_4] \rightarrow [E'_2,\quad E'_3,\quad E'_4,\quad E'_5\cdot E'_1].\]
Therefore we must re-order the cells $\theta$ and $\sigma$ in the
new signatures $A'$, $B'$, $C'$, $D'$, $E'$, which became
particular signatures: see Table 3.

\begin{center}Table 3: Particular signatures\end{center}
\[\ba{|c||c|c|c|c|c|}\hline &A'&B'&C'&D'&E' \\\hline
\hline \theta_1 &  &1/3 &1/3 &1/3 &
\\\hline \theta_2 & & &1/3 &1/3 &1/3
\\\hline \theta_3 &1/3 & & &1/3 &1/3
\\\hline \theta_4 &1/3 &1/3 & & &1/3
\\\hline \theta_5 &1/2 &1/2 &1/2 & &
\\\hline \sigma_{12} &0 & &0 &0 &
\\\hline \sigma_{23} & &0 & &0 &0
\\\hline \sigma_{34} &1/2 & &1/2 & &1/2
\\\hline \sigma_{45} &1/3 &1/3 & &1/3 &
\\\hline \sigma_{51} & &1/3 &1/3 & &1/3
\\\hline \sigma_{13} & & & &2/3 &
\\\hline \sigma_{24} & & & & &1/3
\\\hline \sigma_{35} &1/3 & & & &
\\\hline \sigma_{41} & &1/2 & & &
\\\hline \sigma_{52} & & &2/3 & &
\\\hline \sigma_{134} & & &2/3 & &
\\\hline \sigma_{245} & & & &2/3 &
\\\hline \sigma_{351} & & & & &1/3
\\\hline \sigma_{412} &1/3 & & & &
\\\hline \sigma_{523} & &1/2 & & &
\\\hline \ea\]

Now we must merge these five particular signatures. We will call the resulting signature by $F$.
The merging process is possible because in each row there are
 the same values: see Table 4.

\begin{center}Table 4: Merging of the particular signatures\end{center}
\[\ba{|c||c|c|c|c|c||c|}\hline &A'&B'&C'&D'&E'&F \\\hline
\hline \theta_1 &  &1/3 &1/3 &1/3 & &1/3
\\\hline \theta_2 & & &1/3 &1/3 &1/3&1/3
\\\hline \theta_3 &1/3 & & &1/3 &1/3&1/3
\\\hline \theta_4 &1/3 &1/3 & & &1/3&1/3
\\\hline \theta_5 &1/2 &1/2 &1/2 & &&1/2
\\\hline \sigma_{12} &0 & &0 &0 &&0
\\\hline \sigma_{23} & &0 & &0 &0&0
\\\hline \sigma_{34} &1/2 & &1/2 & &1/2&1/2
\\\hline \sigma_{45} &1/3 &1/3 & &1/3 &&1/3
\\\hline \sigma_{51} & &1/3 &1/3 & &1/3&1/3
\\\hline \sigma_{13} & & & &2/3 &&2/3
\\\hline \sigma_{24} & & & & &1/3&1/3
\\\hline \sigma_{35} &1/3 & & & &&1/3
\\\hline \sigma_{41} & &1/2 & & &&1/2
\\\hline \sigma_{52} & & &2/3 & &&2/3
\\\hline \sigma_{134} & & &2/3 & &&2/3
\\\hline \sigma_{245} & & & &2/3 &&2/3
\\\hline \sigma_{351} & & & & &1/3&1/3
\\\hline \sigma_{412} &1/3 & & & &&1/3
\\\hline \sigma_{523} & &1/2 & & &&1/2
\\\hline \ea\]

The reason for using five 4-signatures, not two of them, for
constructing of 5-signature,
 is that some cells like $\sigma_{13}$ or
$\sigma_{135}$ can be obtained from only one of the signatures
$A,\,B,\,C,\,D,\,E$. Due to this if we do not use all five of the
4-signatures we will not be able to construct the complete
5-signature.

Therefore we get the 5-signature $F$ and, due to the Lemma
\ref{lemma_signature}, it can generate only one tuple of
matrices. However, the existence of this one tuple is not guaranteed by the lemma. Neither it is guaranteed  that this tuple generates a finite orbit.

Now we repeat the same procedure, starting from 4-signatures with
parameters. These signatures are taken from Table 1, possibly
transformed by the symmetry group (\ref{symgroup1}),
(\ref{symgroup2}), (\ref{symgroup3}): $A$ and $B$ are taken from
the orbit number 2 in Table 1, $C$ -- from the orbit 15, $D$ --
from the orbit 1, and $E$ -- from the orbit 5. Some of these
signatures are the members of orbits that are not represented in
the List since the List represents only one member of each orbit.
So we get the Table 5.

\pagebreak

\begin{center}Table 5: 4-signatures with parameters\end{center}
\[\ba{|c||c|c|c|c|c|}\hline &A&B&C&D&E \\\hline
\hline \theta_1 & 0 &a &1/3 &f &2 u+1
\\\hline \theta_2 &x &0 &1/3 &g &u
\\\hline \theta_3 &y &b &1/2 &h &1/3
\\\hline \theta_4 &z &c &1/2 &f+g+h &u
\\\hline \sigma_{12} &x &a &0 &f+g &3 u+1
\\\hline \sigma_{23} &z &b &1/3 &g+h &1/2
\\\hline \sigma_{13} &y &c &2/3 &f+h &1/3
\\\hline \sigma_{24} &y &c &2/3 &f+h &1/3
\\\hline \ea\]

Now we perform  an {\it induction} and obtain the following five
particular 5-signatures: see Table 6.

\begin{center}Table 6: Particular signatures with parameters\end{center}
\[\ba{|c||c|c|c|c|c|}\hline &A'&B'&C'&D'&E' \\\hline
\hline \theta_1 &  &a &1/3 &f &
\\\hline \theta_2 & & &1/3 &g &u
\\\hline \theta_3 &x & & &h &1/3
\\\hline \theta_4 &y &b & & &u
\\\hline \theta_5 &z &c &1/2 & &
\\\hline \sigma_{12} &0 & & 0 &f+g &
\\\hline \sigma_{23} & &0 & &g+h &3 u + 1
\\\hline \sigma_{34} &z & &1/2 & &1/2
\\\hline \sigma_{45} &x &a & &f+g+h &
\\\hline \sigma_{51} & &b &1/3 & &2 u+1
\\\hline \sigma_{13} & & & & f+h &
\\\hline \sigma_{24} & & & & &1/3
\\\hline \sigma_{35} &y & & & &
\\\hline \sigma_{41} & &c & & &
\\\hline \sigma_{52} & & &2/3 & &
\\\hline \sigma_{134} & & &2/3 & &
\\\hline \sigma_{245} & & & &f+h &
\\\hline \sigma_{351} & & & & &1/3
\\\hline \sigma_{412} &y & & & &
\\\hline \sigma_{523} & &c & & &
\\\hline \ea\]

For providing of equivalence of the  values in every row  modulo 2 and up to
sign, we have a finite number of needed relations among the parameters
which are also defined modulo 2.

We choose the same set of relations  as in the previous chapter:

$a = -x, \quad b = -y, \quad c = -z, \quad x = 1/3, \quad y = 1/3,
\quad z = 1/2,$

$f = h = 1/3, \quad g = -1/3, \quad u = 1/3$.

Then we can perform a merging process  of particular signatures
and obtain the signature $F$: see Table 7.

\pagebreak

\begin{center}Table 7: Merging with parameters\end{center}
\[\ba{|c||c|c|c|c|c||c|}\hline &A'&B'&C'&D'&E'&F \\\hline
\hline \theta_1 &  &-1/3 &1/3 &1/3 &&1/3
\\\hline \theta_2 & & &1/3 &-1/3 &1/3&1/3
\\\hline \theta_3 &1/3 & & &1/3 &1/3&1/3
\\\hline \theta_4 &1/3 &-1/3 & & &1/3&1/3
\\\hline \theta_5 &1/2 &-1/2 &1/2 & &&1/2
\\\hline \sigma_{12} &0 & & 0 &0 &&0
\\\hline \sigma_{23} & &0 & &0 &2&0
\\\hline \sigma_{34} &1/2 & &1/2 & &1/2&1/2
\\\hline \sigma_{45} &1/3 &-1/3 & &1/3 &&1/3
\\\hline \sigma_{51} & &-1/3 &1/3 & &5/3&1/3
\\\hline \sigma_{13} & & & & 2/3 &&2/3
\\\hline \sigma_{24} & & & & &1/3&1/3
\\\hline \sigma_{35} &1/3 & & & &&1/3
\\\hline \sigma_{41} & &-1/2 & & &&1/2
\\\hline \sigma_{52} & & &2/3 & &&2/3
\\\hline \sigma_{134} & & &2/3 & &&2/3
\\\hline \sigma_{245} & & & &2/3 &&2/3
\\\hline \sigma_{351} & & & & &1/3&1/3
\\\hline \sigma_{412} &1/3 & & & &&1/3
\\\hline \sigma_{523} & &-1/2 & & &&1/2
\\\hline \ea\]

\section{Constructing algorithm}

In order to get the list of finite orbits of ${\mathcal M}^{(5)}$'s,
we will need the following stages:

\textbf {Stage 1.} We take  a list of signatures of ${\mathcal M_F}^{(4)}$
(including the signatures which correspond to triangular tuples).

We make this list to be finite using free parameters in some
signatures. Further  we allow the next form of cells of signatures
with parameters: each cell can be equal to linear combination of
several parameters,  taken with integer coefficient, and the free
term is a rational number with a denominator $2520$ and determined
modulo 2.  Each parameter is also determined modulo 2.

Moreover, if there is a possibility of equivalent linear
redefinition of the set of parameters with another set such that
Jacobian  of transformation  of parameters is more than $1$ by
absolute value, then we do such the redefinition. E.g. if we have
$\theta_1 = x+y$, $\theta_2 = x - y$ and $\theta_3 = 2 x$, we
 replace it by $\theta_1 = z$, $\theta_2 = w$ and $\theta_3 =
z+w$. The condition of non-existence  of such redefinition will be
called {\it minimal Jacobian condition}.

Each tuple ${\mathcal M_F}^{(4)}$ can be reproducted in 64 copies, using the symmetry
transformations (\ref{symgroup1}), (\ref{symgroup2}) and
(\ref{symgroup3}).

\textbf {Stage 2.} We make all possible 5-signatures, combining the
4-signatures in different ways, as in the example from the previous
chapter. The set of such 5-signatures will be called ${\mathcal S}_5^C$.

Let us construct the signature, which we call ${\mathcal S}$,
meaning that it is the signature of a ${\mathcal M}^{(5)}$, which
consists of $M_1,\,M_2,\,M_3,\,M_4,\,M_5$ monodromy matrices.

First, we take any ${\mathcal M_F}^{(4)}$, defining it as $M_1
M_2,\,M_3,\,M_4,\,M_5$. Knowing the signature of the tuple ${\mathcal
M_F}^{(4)}$, we will have a particular signature of ${\mathcal M}^{(5)}$'s,
containing eight numbers: $\theta_3$, $\theta_4$, $\theta_5$,
$\sigma_{12}$, $\sigma_{34}$, $\sigma_{45}$, $\sigma_{35}$ and
$\sigma_{124}$.

Let us take another ${\mathcal M_F}^{(4)}$, defining it as
$M_1,\,M_2 M_3,\,M_4,\,M_5$, and renaming the eight numbers of its
signature into eight numbers of particular signature of ${\mathcal
M}_5$: $\theta_1$, $\theta_4$, $\theta_5$, $\sigma_{23}$,
$\sigma_{45}$, $\sigma_{15}$, $\sigma_{14}$ and $\sigma_{235}$.
Merging these two particular signatures of the same ${\mathcal
M}_5$ into one particular signature, we demand the coincidence of
two values of $\theta_4$, two of $\theta_5$ and two values of
$\sigma_{45}$. As a result we obtain a particular signature of
${\mathcal M}^{(5)}$ with 13 defined cells.

If the particular signatures that are merging contain parameters, we must
do the following procedure:

For each cell  defined in both particular signatures and if at
least in one of these cells contains parameters, we must consider
two possibilities: 1) the values in these cells  coincide exactly
modulo 2, 2) the values coincide with a change of sign, also
modulo 2. In total, we have two in the power number of such the
cells possibilities.

For each of these possibilities we have a system of linear
equations. To  each cell corresponds one equation, and it is an
equation defined modulo 2, in which variables are also defined
modulo 2, all coefficients are integers, and the free term is a
multiple of $1/2520$.

We solve this system by iterations:

\textbf {Solving of the system}

{\it Step 0.} On each step, we find the smallest in absolute value
non-zero coefficient in the system. Let us call this coefficient
$k_{EV}$. The equation in which it appears will be called $E$, and
the variable by which it appears --- $V$.

{\it Step 1.} If $k_{EV} = \pm 1$, we use the equation $E$ to
determine $V$ via another variables. Then we come back to Step 0,
but with smaller number of variables. Otherwise we  go to Step 2.

{\it Step 2.} If all the coefficients are zero, then the process of solving of
the system is almost finished. We check the free terms. If at
least one  free term is nonzero  modulo 2, then this case produces no solutions. If all equations are satisfied, then  the system is solved and the task is performed.

{\it Step 3.} If $|k_{EV}| \geq 2$, we look for the
coefficient by the variable $V$ in  other equations, which is
bigger or equal to $k_{EV}$ by absolute value. If it exists, we add to
each such equation the equation $E$ with such a factor that coefficient by
$V$ becomes less than $k_{EV}$ in absolute value. Then we come
back to Step 0, having either smaller  minimal coefficient or
smaller number of nonzero coefficients. Otherwise we go to Step 4.

{\it Step 4.} We check if there is a coefficient in $E$ which is
bigger or equal to $k_{EV}$ by absolute value. If it exists, we
redefine the variable $V$ by adding to it other variables with
such factors that  there remains no coefficient bigger or equal
than $k_{EV}$ in the equation $E$. Then we come back to Step 0,
having either smaller
 minimal coefficient or smaller number of nonzero coefficients.
Otherwise we go to Step 5.

{\it Step 5.} Here we have $|k_{EV}|\geq 2$, all other
coefficients by $V$ are zero, and all other coefficients in the
equation $E$ are zero too. We control that $|k_{EV}|$ is a divisor
of $5040$ and the free term of $E$ is a multiple of $|k_{EV}| /
2520$. If one of these conditions is violated, we treat this as a
mistake of the algorithm. In fact, it never happens. Therefore we
choose an integer value $m$ from the interval $[0,\,|k_{EV}|-1]$
(yet another branching of possibilities), divide the equation $E$
by $k_{EV}$ and add to its free term $2 m / k_{EV}$. We have to do
this because the equation is  defined modulo $2$. And we come back
to Step 0. $\square$

Solving the system, we perform the operations with variables
(see Steps 1, 4, 5) simultaneously  with parameters in signatures.
When the system is solved, we get new particular signatures
with new parameters (number of parameters can increase, decrease
or even vanish).

In order to provide the {\it minimal Jacobian condition}, we
transform the parameters in obtained particular signature with
Gauss-like transformation.

Note that this procedure can give several results in case of
particular signatures with parameters.

Next, we repeat this procedure thrice, merging the particular
signature with particular signatures obtained by induction of
${\mathcal M_F}^{(4)}$'s, defined as
 $\{M_1,\,M_2,$ $M_3\cdot M_4,\,M_5\}$, $\{M_1,\,M_2,\,M_3,\,M_4\cdot M_5\}$ and
$\{M_2,\,M_3,\,M_4,\,M_5\cdot M_1\}$, in order to fill in all cells and obtain the
complete signature ${\mathcal S}_5$, which is considered to be a
signature of ${\mathcal M}^{(5)}$.

Using this procedure with all 4-signatures from Table 1 (which are
the signatures of all ${\mathcal M_F}^{(4)}$'s), we get a list of
${\mathcal S}_5$'s which we call ${\mathcal S^C}_5$. We know that
the signature of every ${\mathcal M_F}^{(5)}$ belongs to this
list.

Naively, we would check each signature ${\mathcal S^C}_5$ for {\it
consistency}. The {\it inconsistent} ones  we remove from the list.
And for each consistent signature we  reconstruct  ${\mathcal M}^{(5)}$ as a
tuple of matrices and check whether it generates a finite orbit.

But here there is a problem: the check of the {\it consistency} with the help of the
computer program can not be done exactly and, moreover,  for the tuple ${\mathcal
M}_5$'s containing free parameters this task is impossible to perform with the help of  computer algebra.

That is why we will proceed in another way.

\textbf {Stage 3.}

The straightforward way to  proceed would be the following sequence of steps:

Each signature from the list ${\mathcal S^C}_5$ we should check
for consistency, and --- if it is consistent -- to reconstruct it
into ${\mathcal M}^{(5)}$. Then, using these ${\mathcal
M}^{(5)}$'s we should form the list and call it ${\mathcal
M^C}_5$. Then on each element $M$ of ${\mathcal M^C}_5$ we should
act by ten braid group actions (see (\ref{braida1}) and
(\ref{braida2})) and check whether all the results of these
actions also belong to ${\mathcal M^C}_5$. If they do  not, than
we would be sure that $M$ generates an infinite orbit, and exclude
it from the list ${\mathcal M^C}_5$. Then we should  repeat this
check until there remains nothing to exclude. After this, we would
be sure that ${\mathcal M^C}_5$ contains only finite orbits, or
possibly infinite orbits written by finite number of elements with
parameters, so that during continuation of the orbit the
parameters transform under an infinite group action.

But our wish is to avoid this complicated procedure of checking
the consistency and reconstruction of tuples of matrices.

That's why we will use another approach. Starting from the list
${\mathcal S^C}_5$, we will treat it, directly considering the
signatures and removing unwanted elements from this list, and the
rest elements uniting into the orbits.

It is possible to determine the braid group actions onto the
signature, and the result will be an {\it incomplete signature}.

Due to the definition of the incomplete signature, there are ten
types of incomplete signatures, each containing sixteen cells: all
five $\theta$'s, the $\sigma_{12}$, $\sigma_{23}$, $\sigma_{34}$,
$\sigma_{45}$, $\sigma_{15}$ and six more $\sigma$'s. Each type of
the incomplete signatures is obtained from complete signatures by
a specific braid group action. In more details,  if we act on any
signature by braid group action ${\mathcal B}_{a,b}$, then we get
an incomplete signature, in which  $\sigma$'s containing  index
$a$ and not containing $b$ will be undefined.

If we have ${\mathcal M}^{(5)}$ which will be denoted $M$, and  $S =
\{\theta_1...\}$ is the signature of $M$, then the signature $S' =
\{\theta'_1...\}$ of the result of braid group action ${\mathcal
B}_{k,k+1} M$  will be

\[\ba{l}
\theta'_m = \theta_m,\quad m\neq k,\,k+1;\\
\theta'_k = \theta_{k+1},\quad \theta'_{k+1} = \theta_{k};\\
\sigma'_{a,b...} = \sigma_{a,b...},\quad k,\,k+1 \notin
\{a,b...\};\\
\sigma'_{...k,k+1...} = \sigma_{...k,k+1...};\\
\sigma'_{...k...} = \,\,?;\\
\sigma'_{...k+1...} = \sigma_{...k...}.\\
\ea\]

The  symbol $?$ means that there is no simple way to calculate this cell,
and we leave it undetermined.

Similarly, the signature $S'' = \{\theta''_1...\}$ of ${\mathcal
B}_{k+1,k} M$ will be

\[\ba{l}
\theta''_m = \theta_m,\quad m\neq k,\,k+1;\\
\theta''_k = \theta_{k+1},\quad \theta''_{k+1} = \theta_{k};\\
\sigma''_{a,b...} = \sigma_{a,b...},\quad k,\,k+1 \notin
\{a,b...\};\\
\sigma''_{...k,k+1...} = \sigma_{...k,k+1...};\\
\sigma''_{...k...} = \sigma_{...k+1...};\\
\sigma''_{...k+1...} = \,\,?.\\
\ea\]

We remind that all indices here  are modulo $n$.

Therefore our plan is the following:

1. We take any signature from ${\mathcal S^C}_5$ and call it $S$.

2. We act on $S$ by all $2\cdot n = 10$ braid group actions.

3. The result of each braid group action on $S$ is an incomplete
signature which we call $S'$.

4. If $S$ does not contain independent parameters -- we look in
${\mathcal S^C}_5$ for any signature which can be merged with
$S'$. If there is no such the signature, we exclude the $S$ from ${\mathcal S^C}_5$.

5. If $S$ contains independent parameters, and so does $S'$  -- we look (in ${\mathcal S^C}_5$)
 for any signature which can be merged with
$S'$ without imposing  conditions on parameters in $S'$.

6. If $S$ contains independent parameters, and the step 5 fails,
we look in ${\mathcal S^C}_5$ for any signature, let us call it $S^*$,
which can be merged with $S'$ after the imposing  conditions on
parameters in $S'$. For each such $S^*$ (it can be one, more than
one or none) we make a copy of $S$ after imposing the same
conditions on its parameters and add it to ${\mathcal S^C}_5$.  After this we
exclude $S$ from ${\mathcal S^C}_5$.

7. We repeat this procedure for all members of ${\mathcal S^C}_5$
until it remains nothing to exclude.$\square$

Note that some signatures can be inconsistent, but since the braid
group actions are natural only for consistent signatures -- the
inconsistent signatures are likely to be excluded by this
procedure. However, some inconsistent signatures can remain in the
list.

Note that we could write down the same procedure only for
signatures without parameters,  avoiding the steps 5 and 6, but it
would be an infinite procedure with infinite list of signatures. Due to the steps 5
and 6 we obtain the same result by a finite sequence of
steps.

To illustrate these procedure, we provide three examples (all
gathered in the Table 8).

First example is a 5-signature $A$ without parameters, which is
the signature of ${\mathcal M_F}^{(5)}$ from (\ref{example}). We
perform the braid group action ${\mathcal B}_{3,2}$ getting an
incomplete signature $A'$ and find in ${\mathcal S^C}_5$ the
signature $A^*$ which can be merged with $A'$, so the Step 4
succeeds.

Second example is the signature $B$ such that $B' = {\mathcal
B}_{3,2} B$ can not be merged with any signature in ${\mathcal
S^C}_5$.

In the third example we start from the signature $C$ with
parameters. For it, we get an incomplete signature $C' = {\mathcal
B}_{3,2} C$. In order to merge $C'$ with any member of ${\mathcal
S^C}_5$, which will be called $C^*$, we take an element of
${\mathcal S^C}_5$, same as $C$, of course with other notations
for its parameters: $a$, $b$, $c$ replacing $x$, $y$, $z$.
Further, during merging of $C'$ and $C^*$ we require some
conditions on the parameters, and one of the possibilities for
such conditions is $a = b = x = y = 0$,  $c = z = 1/2$. The result
of merging is called $C'\wedge C^*$ and the copy of $C$ which is
added to ${\mathcal S^C}_5$ after excluding $C$ is called
$C_\times$.

\pagebreak

\begin{center}Table 8: Braiding of 5-signatures\end{center}
\[ \ba{|c||c|c|c||c|c||c|c|c|c|c|}\hline &A&A'&A^* &B&B' &C&C'&C^*
& C'\wedge C^* & C_\times
\\\hline
\hline \theta_1       &1/3  &1/3  &1/3  &1/2  &1/2  &y    &y    &b
&0 &0
\\\hline \theta_2     &1/3  &1/3  &1/3  &1/2  &1/2  &z    &1/2  &c  &1/2
&1/2
\\\hline \theta_3     &1/3  &1/3  &1/3  &1/2  &1/2  &1/2  &z    &1/2  &1/2
&1/2
\\\hline \theta_4     &1/3  &1/3  &1/3  &1/2  &1/2  &x    &x    &a    &0
&0
\\\hline \theta_5     &1/2  &1/2  &1/2  &1/2  &1/2  &1/2  &1/2  &1/2  &1/2
&1/2
\\\hline \sigma_{12}  &0    &2/3  &2/3  &1/3  &3/5  &y+z  &1/2  &b+c  &1/2
&1/2
\\\hline \sigma_{23}  &0    &0    &0    &1/5  &1/5  &1/2  &1/2  &1/2  &1/2
&1/2
\\\hline \sigma_{34}  &1/2  &1/3  &1/3  &1/5  &4/5  &1/2  &x-z  &1/2  &1/2
&1/2
\\\hline \sigma_{45}  &1/3  &1/3  &1/3  &1/3  &1/3  &1/2  &1/2  &1/2  &1/2
&1/2
\\\hline \sigma_{51}  &1/3  &1/3  &1/3  &2/5  &2/5  &1/2  &1/2  &1/2  &1/2
&1/2
\\\hline \sigma_{13}  &2/3  &      &0    &3/5  &  &1/2  &     &1/2  &1/2
&1/2
\\\hline \sigma_{24}  &1/3  &1/2   &1/2    &2/3  &1/5  &x+z  &1/2  &a+c  &1/2
&1/2
\\\hline \sigma_{35}  &1/3  &      &2/3  &3/5  &  &x+y+z&     &a+b+c  &1/2
&1/2
\\\hline \sigma_{41}  &1/2  &1/2  &1/2  &1/2  &1/2  &x+y  &x+y  &a+b    &0
&0
\\\hline \sigma_{52}  &2/3  &1/3  &1/3  &1/2  &3/5  &1/2  &x+y+z&1/2  &1/2
&1/2
\\\hline \sigma_{134} &2/3  &     &1/3  &3/5  &  &1/2  &     &1/2  &1/2
&1/2
\\\hline \sigma_{245} &2/3  &0    &0    &2/3  &1/3  &1/2  &y+z  &1/2  &1/2
&1/2
\\\hline \sigma_{351} &1/3  &     &1/2  &4/5  &  &x-z  &     &a-c  &1/2
&1/2
\\\hline \sigma_{412} &1/3  &2/3  &2/3  &2/3  &3/5  &x-y-z&1/2  &a-b-c  &1/2
&1/2
\\\hline \sigma_{523} &1/2  &1/2  &1/2  &3/5  &3/5  &x-y  &x-y  &a-b  &0&0
\\\hline \ea\]

\textbf{Stage 4.} After all exclusions, the list ${\mathcal S^C}_5$
will be renamed as ${\mathcal S^F}_5$.

Therefore ${\mathcal S^F}_5$ is the list of signatures, closed under the braid
group actions. That is why ${\mathcal S^F}_5$ is split under braid group actions into a number of
pieces. But these pieces still are not the orbits.

Naively, we would do two steps to finally construct the orbits:

1. First, we must check consistency of all these signatures. If one
signature is inconsistent, then all signatures from the piece,
associated with it by braid group actions, are inconsistent too.
After excluding inconsistent signatures, only the consistent ones
remain.

Now we can transform this list of signatures into list of tuples
${\mathcal M}^{(5)}$'s.

2. For each tuple ${\mathcal M}^{(5)}$ from this list we have to
construct an orbit to make sure that it is a finite orbit. If
${\mathcal M}^{(5)}$ contains no parameters, the orbit must be
finite because the number of ${\mathcal M}^{(5)}$'s without
parameters in the list is finite. As for any ${\mathcal M}^{(5)}$
with parameters, if the procedure of construction of the orbit
does not terminate for too many steps, we will try to write this
orbit with a finite set of elements and introduce a group of
transformation of the parameters.

In fact, we are sure that every tuple ${\mathcal M}^{(5)}$ with one
parameter generates a finite orbit, because the order of the group of
transformations of one parameter cannot be bigger than $10080$:
transformations of the parameters must be linear, with integer
coefficient, invertible, that's why the coefficient can be $\pm 1$
only, and the free term must be a multiple of $1/2520$ and defined
by modulo 2.

Therefore, the decision about "construction of the orbit does not
terminate for too many steps" we must do only for the orbits with two
or more parameters.

In order to simplify calculations, we perform another procedure with the list ${\mathcal
S^F}_5$ to obtain the same result:

We construct several classes of the tuples ${\mathcal M_F}^{(5)}$'s, which can be
described simply as follows:

1. All ${\mathcal M_F}^{(4)}$'s plus the unit matrix.

2-6. All matrices in a ${\mathcal M_F}^{(5)}$ belong to a finite
subgroup of  $SU(2)$ group. There are five such subgroups:

2. Cyclical group. All the matrices are diagonal.

3. Dihedral group. Every matrix is either diagonal or its
diagonal elements are zeros.

4. Tetrahedral group.

5. Octahedral group.

6. Icosahedral group.

For all these tuples ${\mathcal M_F}^{(5)}$ we make 5-signatures and find the
same signatures in list ${\mathcal S^F}_5$. For the types 1, 4, 5, 6
we can do it straightforwardly. As for types 2, 3 there exists an
infinite number of cyclical and dihedral groups, but belonging of
the signature to one of these groups can be checked by simple
calculations.

All this can be done using a specially designed computer program.

The members of the list ${\mathcal S^F}_5$ which are not of the
types 1, 2, 3, 4, 5, 6, we will consider manually.

\section{Computer realization}

At this stage, the classification is formulated in such a way that it can be
carried out with a specially designed program.

The computer program for the algorithm from the previous chapter was
written in the C++ language and ran on a personal computer for about
12 hours.

To have confidence in accuracy of the calculations, we restricted
the use of floating point numbers: double and complex variables
are used only for constructing the list of 4-signatures and for
constructing of tetrahedral, octahedral and icosahedral groups
(for which the correct results are well known), for approximate
checking of consistency of signatures without parameters (which
result is only informative and does not influence further
calculations) and also for the visualization of the progress bar.

In the program there is a variable called "errorcode". Normally it
is zero, but in every abnormal situation it is assigned the code
of the situation, and can never become zero again. The fact that
this variable remains zero till the program finishes makes us
believe that the program works correctly.

The values of the coefficients of the parameters in signature cells
never exceed $4$, so we are not afraid of arithmetical overflow.

Free terms in the cells are standard fractions with
denominator $2520$ and numerator an integer value which does not
exceed $2520$ by absolute value.

The source code for this program is available by request.

\section{The result of computer calculations}

The result of computer calculations is the list ${\mathcal
S^F}_5$ which consists of $231$ orbits of signatures (we call an {\it orbit of
signatures} the subset of ${\mathcal S^F}_5$ connected with braid
group actions).


From these $231$, $128$ were recognized as constructed from
${\mathcal M}^{(4)}$ by addition of unit matrix.

And from the remaining $103$, $3$, $19$ and $71$ were recognized
respectively as tetrahedral, octahedral and icosahedral type (in
Table 9, see Sect. 10 they are numbered respectively as $9-11$,
$12-30$ and $31-101$).

There remain ten orbits of signatures: one with four parameters,
two with three parameters, two with one parameter and five without
parameters.

They are presented in the Table 10, where each orbit is
represented by one signature.

\begin{center}Table 10: Ten exceptional orbits\end{center}
\[\ba{|c|c||c|c|c|c|c||c|c|c|c|c|c|c|} \hline
\sharp & length &\theta_1 &\theta_2 &\theta_3 &\theta_4 &\theta_5 &\sigma_{12} &\sigma_{23} &\sigma_{34} &\sigma_{45} &\sigma_{51} &\sigma_{13} &\sigma_{24}\\
\hline 0& &x & y & z & w & \ba{c}x\!+\!y\!+\\z\!+\!w\ea
& x+y & y+z & z+w & \ba{c}x+y\\+z\ea & \ba{c}y\!+\\z\!+\!w\ea & x\!+\!z & y+w\\
\hline 1 & 4 &1/2 & x & y & z & 1/2 & 1/2 & x+y & y+z & 1/2 &
 \!x\!\!+\!\!y\!\!+\!\!z\!  & 1/2 & x+z\\
\hline 2 &
&1/2 & 1/2 & x & 1/2 & 1/2 & x+y & 1/2 & 1/2 & y & z & 1/2 & x+z\\
\hline 3 &
9 &x & x & x & x & 2/3 & 2 x & 2 x& 1/3 & 3 x & 1/2 & 2 x & 1/3\\
\hline 4&12 &x & x & 2 x\!\!+\!\!1\! &
x & x & 1/3 & 1/2 & 1/2 & 1/3 & 2 x & 3 x\!\!+\!\!1\! & 2 x\\
\hline 5&105 &2/7 & 2/7 & 2/7 & 2/7 & 2/7 & 1/3 & 1/7 & 1/7 & 1/7
& 1/3 & 1/3 & 1/2\\ \hline 6&105 &4/7 & 4/7 & 4/7 & 4/7 & 4/7 &
1/3 & 5/7 & 5/7 & 5/7 & 1/3 & 1/3 & 1/2\\\hline 7&105 &6/7 & 6/7 &
6/7 &
6/7 & 6/7 & 1/3 & 3/7 & 3/7 & 3/7 & 1/3 & 1/3 & 1/2\\
\hline 8&192 &1 & 1 & 1 & 1 & 1 & 1/3 & 1/3 & 1/3 & 1/3 & 1/3 &
1/5 & 1/5\\ \hline A&1 &0 & 0 & 0 & 0 & 0 & 1 & 1 & 1 & 1 & 1 & 1
& 1\\ \hline \ea\]

The orbit with number $0$ with four parameters cooresponds to the
{\it triangular} ${\mathcal M}^{(5)}$'s, thus it is outside the
scope this paper and will be not considered below. For this type
of orbits we can say also that it includes the case when all
matrices belong to a cyclical subgroup of $SU(2)$ (i.e. when all
matrices are diagonal), but it also includes other subtypes: when
all matrices cannot be diagonalized simultaneously, but are lower
triangular and some of them have nonzero $[2,1]$ element.
Nevertheless, the signatures of ${\mathcal M}^{(n)}$'s of orbits
belonging to these other subtypes are the same as signatures of
${\mathcal M}^{(n)}$ where all matrices belong to a cyclical
group.

The tuple ${\mathcal M}^{(5)}$ for the orbit 0 is constructed in (\ref{lowertriangular}):

\[M_1 = \left(\ba{cc}X&0\\V_x&X^{-1}\ea\right),\quad M_2 = \left(\ba{cc}Y&0\\V_y&Y^{-1}\ea\right),\quad M_3 =
\left(\ba{cc}Z&0\\V_z&Z^{-1}\ea\right),\]\be\label{lowertriangular} M_4 =
\left(\ba{cc}W&0\\V_w&W^{-1}\ea\right),\quad M_5 =
\left(\ba{cc}(X\,Y\,Z\,W)^{-1}&0\\V_5&X\,Y\,Z\,W\ea\right)\ee

with different constraints on the parameters $X,\,Y,\,Z,\,W$ and the off-diagonal elements
$V_x$, $V_y$, $V_z$, $V_w$, $V_5$.

For orbits $1-8$ in the list the tuples of matrices were constructed
explicitly, using notations $X = \exp(i\pi \,x)$, $Y = \exp(i\pi
\,y)$, $Z = \exp(i\pi \,z)$.

Orbit 1:
\[M_1 = \left(\ba{cc}0&1\\-1&0\ea\right),\quad M_2 = \left(\ba{cc}X&0\\0&X^{-1}\ea\right),\quad M_3 = \left(\ba{cc}Y&0\\0&Y^{-1}\ea\right),
\]\be\label{3letter4} M_4 = \left(\ba{cc}Z&0\\0&Z^{-1}\ea\right),\quad
M_5 = \left(\ba{cc}0&-(X\, Y\, Z)^{-1}\\X\, Y\, Z&0\ea\right).\ee
The length of this orbit is 4.

Orbit 2:\[M_1 = \left(\ba{cc}0&1\\-1&0\ea\right),\quad M_2 = \left(\ba{cc}0&-X\, Y\\
(X\, Y)^{-1}& 0 \ea\right), \quad M_3 =
\left(\ba{cc}X&0\\0&X^{-1}\ea\right),
\]\be\label{3letterbig} M_4 = \left(\ba{cc}0&Y Z^{-1}\\-Y^{-1}
Z&0\ea\right),\quad M_5 =
\left(\ba{cc}0&-Z^{-1}\\Z&0\ea\right),\ee

where $x,\,y,\,z$ must be rational numbers. The length of this
orbit is
\[ \frac{u^2 \, v^2 \prod_{p\geq 3} (1-p^{-2})}{1 + \delta_{u,1}(1-\delta_{v,1})},\]
where $u$ is the denominator of $x$, $v$ is the smallest
common denominator of values $(u\,y)$ and $(u\, z)$, and $p$ is an odd
prime divisor of $v$; the denominator of this equation is $2$ for
$u=1$ and $v\geq 2$, and $1$ otherwise.

In orbits 1 and 2 all matrices belong to the dihedral group.

Orbit 3:
\[ M_1 = \left(\ba{ll}X&0\\1&X^{-1}\ea\right), \quad
M_2 = \left(\ba{ll}X&0\\1&X^{-1}\ea\right), \quad M_3 =
\left(\ba{ll}X&0\\1&X^{-1}\ea\right), \] \be\label{1letter9} M_4 =
\left(\ba{ll}X^{-1}&-1\\0&X\ea\right),\quad M_5 =
\left(\ba{ll}-1-X^2&X^3\\-X-X^{-1}-X^{-3}&X^2\ea\right).\ee

Orbit 4: \[M_1 = \left(\ba{ll}X&0\\-1&X^{-1}\ea\right),\quad M_2 =
\left(\ba{ll}X^{-1}&1\\0&X\ea\right),\quad M_3 =
\left(\ba{ll}-X^2&0\\X+X^{-1}&-X^{-2}\ea\right),\]
\be\label{1letter12} M_4 =
\left(\ba{ll}X^{-1}&1\\0&X\ea\right),\quad M_5 =
\left(\ba{ll}X&0\\-1&X^{-1}\ea\right).\ee

Orbits 5, 6, 7 can be written in the same form: \[M_1 =
\left(\ba{ll}s&0\\0& s^6\ea\right),\]\[M_2 = \left(\ba{ll}
s^6(s-1)^5 (1 + s)^3 (1 + s^2)^2 / 7&
  s^5 (1 + s^2) / 7 \\ s(s-1)^4 (1 + s)^3&
   s^3(1 - s)^5 (1 + s)^3 (1 + s^2)^2 / 7 \ea\right), \]
\be\label{noletter105} M_3 = M_1^2\, M_2 \, M_1^{-2},\quad M_4 =
M_1^{-3}\, M_2 \, M_1^3,\quad M_5 = M_1^{-1} \, M_2\, M_1,\ee

where $s$ is one of the three seventh roots of unity: $s = \exp(2\,i\pi/7)$,  $s
= \exp(4\,i\pi/7)$ or $s = \exp(6\,i\pi/7)$. Replacing $s\rightarrow s^{-1}$ yields the same matrices up to
a common conjugation.

Orbit 8: \[M_1 = \left(\ba{ll}-1& 1\\0& -1 \ea\right),\quad M_2 =
\left(\ba{ll}-1& 0\\-1& -1\ea\right),\quad M_3 =
\left(\ba{ll}\frac{-1 - \sqrt{5}}{2}&
  1\\\frac{-3 + \sqrt{5}}{2}& \frac{-3 + \sqrt{5}}{2}\ea\right),\]\be\label{noletter192}M_4 = \left(\ba{ll}\frac{1 -
  \sqrt{5}}{2}&
  \frac{3 -\sqrt{5}}{2}\\\frac{-3 + \sqrt{5}}{2}&
\frac{-5 + \sqrt{5}}{2}\ea\right),\quad M_5 =
\left(\ba{ll}\frac{-1 - \sqrt{5}}{2}& \frac{3 -
  \sqrt{5}}{2}\\-1&
\frac{-3 + \sqrt{5}}{2}\ea\right).\ee

The orbit in ${\mathcal S^F}_5$, which is called $A$ in the list,
consists of one element, turns out to be inconsistent: all
$\theta$'s are equal to $0$ and all $\sigma$'s are equal to $1$.
It is easy to try to reconstruct the matrix tuple from it and
confirm that it is impossible. Therefore this signature must be
excluded.

It is not surprising that only one signature turned out to be
inconsistent, because the braid group actions are originally
defined for tuples of matrices. Thus the result of braid
group action on an inconsistent signature can only coincide with another signature
from any list by accident only.

\section{List of signatures of ${\mathcal M_F}^{(5)}$'s}

Here we present the list of signatures of ${\mathcal M_F}^{(5)}$'s. From
this list we omit the ${\mathcal M_F}^{(5)}$'s which are obtained from
${\mathcal M_F}^{(4)}$'s by addition of the unit matrix, and the triangular
${\mathcal M_F}^{(5)}$'s.

For each symmetry class of orbits under (\ref{symgroup1}),
(\ref{symgroup2}) and (\ref{symgroup3}) we present only one
the representative orbit, and for each orbit -- only one element (we call it
start element, though every element of an orbit can be chosen as
the start element).

For each signature, we present not all $20$ cells, but only all
$\theta$'s and seven of the $\sigma$'s, because it is enough to
reconstruct the tuple of matrices.

\pagebreak

\begin{center}Table 9: List of 5-signatures which generate finite orbits\end{center}
\[\ba{|c|c||c|c|c|c|c||c|c|c|c|c|c|c|} \hline
\sharp & length &\theta_1 &\theta_2 &\theta_3 &\theta_4 &\theta_5 &\sigma_{12} &\sigma_{23} &\sigma_{34} &\sigma_{45} &\sigma_{51} &\sigma_{13} &\sigma_{24}\\
\hline
\hline 1 & 4 &1/2 & x & y & z & 1/2 & 1/2 & x+y & y+z
& 1/2 &  x\!+\!y\!+\!z  & 1/2 & x+z\\
\hline 2 &
&1/2 & 1/2 & x & 1/2 & 1/2 & x+y & 1/2 & 1/2 & y & z & 1/2 & x+z\\
\hline 3 &
9 &x & x & x & x & 2/3 & 2 x & 2 x& 1/3 & 3 x & 1/2 & 2 x & 1/3\\
\hline 4&12 &x & x & 2 x\!+\!1 &
x & x & 1/3 & 1/2 & 1/2 & 1/3 & 2 x & 3 x+1 & 2 x\\
\hline 5&105 &2/7 & 2/7 & 2/7 & 2/7 & 2/7 & 1/3 & 1/7 & 1/7 & 1/7
& 1/3 & 1/3 & 1/2\\ \hline 6&105 &4/7 & 4/7 & 4/7 & 4/7 & 4/7 &
1/3 & 5/7 & 5/7 & 5/7 & 1/3 & 1/3 & 1/2\\\hline 7&105 &6/7 & 6/7 &
6/7 &
6/7 & 6/7 & 1/3 & 3/7 & 3/7 & 3/7 & 1/3 & 1/3 & 1/2\\
\hline 8&192 &1 & 1 & 1 & 1 & 1 & 1/3 & 1/3 & 1/3 & 1/3 & 1/3 &
1/5 & 1/5\\ \hline
\hline 9&16 &1/3 & 1/3 & 1/3 & 1/3 & 1/2 & 0 & 0 & 1/2 & 1/3 & 1/3
& 2/3 & 1/3\\ \hline 10&24 &1/3 & 1/3 & 1/3 & 1/2 & 1/2 & 1/3 &
1/3 & 1/3 & 0 & 1/3 & 1/3 & 1/3\\ \hline 11&36 &1/3 & 1/3 & 1/2 &
1/2 & 1/2 & 0 & 1/3 & 1/2 & 1/2 & 1/3 & 2/3 & 1/3\\ \hline
\hline 12&18 &1/4 & 1/4 & 1/3 & 1/2 & 1/2 & 1/3 & 1/4 & 1/3 & 0 &
1/4 & 1/4 & 1/2\\ \hline 13&24 &1/4 & 1/4 & 1/4 & 1/2 & 1/2 & 0 &
1/3 & 1/2 & 1/4 & 1/4 & 1/3 & 1/3\\ \hline 14&24 &1/4 & 1/4 & 1/3
& 1/3 & 1/2 & 0 & 1/4 & 1/2 & 1/3 & 1/4 & 1/2 & 1/4\\ \hline 15&24
&1/4
& 1/4 & 1/3 & 1/3 & 2/3 & 0 & 1/4 & 2/3 & 1/3 & 1/2 & 1/2 & 1/4\\
\hline 16&27 &1/4 & 1/4 & 1/4 & 1/4 & 1/3 & 0 & 0 & 1/3 & 1/4 &
1/4 & 1/2 & 1/3\\ \hline 17&36 &1/4 & 1/3 & 1/2 & 1/2 & 1/2 & 1/4
& 1/4 & 1/4 & 1/2 & 1/2 & 2/3 & 2/3\\ \hline 18&36 &1/4 & 1/4 &
1/4 & 1/3 & 1/2 & 0 & 0 & 1/2 & 1/4 & 1/3 & 1/2 & 1/4\\ \hline
19&40 &1/4 &
1/4 & 1/3 & 1/3 & 1/3 & 0 & 1/4 & 1/3 & 1/3 & 1/4 & 1/2 & 1/4\\
\hline 20&48 &1/4
& 1/4 & 1/2 & 1/2 & 1/2 & 0 & 1/3 & 1/2 & 1/2 & 1/3 & 2/3 & 1/2\\
\hline 21&48 &1/4 & 1/3 & 1/3 & 1/2 & 1/2 & 1/4 & 0 & 1/4 & 1/4 &
1/2 & 1/2 & 3/4\\ \hline 22&64 &1/4 & 1/3 & 1/3 & 1/3 & 1/2 & 1/4
& 0 & 1/3 & 1/4 & 1/3 & 1/2 & 1/2\\ \hline 23&72 &1/3 & 1/2 & 1/2
& 1/2 & 1/2 & 1/4 & 1/4 & 1/4 & 1/2 & 2/3 & 2/3 & 2/3\\ \hline
24&72 &1/4 &
1/4 & 1/3 & 1/2 & 1/2 & 0 & 1/4 & 1/2 & 1/3 & 1/3 & 1/2 & 1/3\\
\hline 25&96 &1/4 & 1/2 & 1/2 & 1/2 & 1/2 & 1/4 & 1/4 & 1/3 & 1/3
& 2/3 & 2/3 & 1/2\\ \hline 26&96 &1/3 & 1/3 & 1/2 & 1/2 & 1/2 & 0
& 1/4 & 1/2 & 1/2 & 1/3 & 3/4 & 1/2\\ \hline 27&120 &1/3 & 1/3 &
1/3 & 1/2 & 1/2 & 0 & 1/3 & 1/2 & 1/3 & 1/4 & 1/2 & 1/4\\ \hline
28&144 &1/4 &
1/3 & 1/2 & 1/2 & 1/2 & 1/4 & 1/4 & 1/3 & 1/3 & 1/3 & 1/2 & 1/2\\
\hline 29&192 &1/2 & 1/2 & 1/2 & 1/2 & 1/2 & 0 & 1/4 & 1/2 & 1/2 &
1/3 & 3/4 & 1/3\\ \hline 30&216 &1/3 & 1/2 & 1/2 & 1/2 & 1/2 & 1/4
& 0 & 1/3 & 1/3 & 1/2 & 3/4 & 2/3\\ \hline
\hline 31&30 &1/5 & 1/5 & 2/5 & 2/5 & 2/5 & 1/5 & 1/5 & 1/3 & 1/3
& 1/3 & 1/2 & 1/2\\ \hline 32&30 &1/5 & 1/5 & 1/5 & 2/5 & 3/5 & 0
& 1/5 & 3/5 & 1/5 & 1/2 & 1/3 & 1/3\\ \hline 33&36 &1/5 & 1/3 &
2/5 & 2/5 & 3/5 & 1/5 & 1/5 & 4/5 & 1/5 & 1/2 & 1/2 & 1/5\\ \hline
34&36 &1/5 &
1/5 & 1/5 & 1/3 & 3/5 & 1/5 & 1/5 & 2/5 & 1/3 & 1/2 & 1/3 & 2/5\\
\hline 35&40 &1/5 & 1/5 & 2/5 & 2/5 & 3/5 & 0 & 1/5 & 3/5 & 2/5 &
1/2 & 3/5 & 1/3\\ \hline 36&40 &1/5 & 1/5 & 1/5 & 2/5 & 2/5 & 1/5
& 1/5 & 1/5 & 0 & 1/3 & 1/5 & 1/2\\ \hline 37&45 &1/5 & 2/5 & 2/5
& 2/5 & 3/5 & 1/3 & 0 & 3/5 & 1/5 & 2/5 & 1/2 & 1/3\\ \hline 38&45
&1/3 &
1/3 & 2/5 & 2/5 & 2/5 & 1/5 & 1/5 & 1/3 & 1/3 & 1/5 & 1/2 & 1/2\\
\hline 39&45 &1/5 & 1/5 & 1/3 & 2/5 & 3/5 & 1/5 & 1/5 & 2/5 & 2/5
& 1/2 & 1/2 & 1/2\\ \hline 40&45 &1/5 & 1/5 & 1/5 & 1/5 & 2/5 & 0
& 1/5 & 2/5 & 1/5 & 1/3 & 1/3 & 1/5\\ \hline 41&45 &1/5 & 1/5 &
1/5 & 1/3 & 2/3 & 1/5 & 1/5 & 1/2 & 1/3 & 1/2 & 1/3 & 1/3\\ \hline
\ea\]

\[\ba{|c|c||c|c|c|c|c||c|c|c|c|c|c|c|} \hline
\sharp& length &\theta_1 &\theta_2 &\theta_3 &\theta_4 &\theta_5 &\sigma_{12} &\sigma_{23} &\sigma_{34} &\sigma_{45} &\sigma_{51} &\sigma_{13} &\sigma_{24}\\
\hline  42&64 &2/5 & 2/5
& 2/5 & 2/5 & 1/2 & 1/3 & 1/3 & 1/3 & 1/5 & 1/5 & 1/3 & 1/3\\
\hline 43&64 &1/5 & 1/5 & 1/5 & 1/5 & 1/2 & 1/5 & 1/5 & 1/3 & 1/3
& 1/3 & 1/3 & 1/5\\ \hline 44&72 &1/5 & 1/5 & 1/3 & 2/5 & 2/5 & 0
& 1/5 & 2/5 & 1/3 & 1/3 & 1/2 & 1/3\\ \hline 45&80 &1/5 & 1/5 &
2/5 & 2/5 & 1/2 & 1/5 & 1/5 & 1/3 & 1/3 & 1/3 & 1/2 & 1/2\\ \hline
46&81 &1/5 &
1/3 & 2/5 & 2/5 & 2/5 & 1/5 & 1/5 & 1/3 & 1/3 & 1/3 & 1/2 & 2/5\\
\hline 47&81 &1/5 & 1/5 & 1/5 & 1/3 & 2/5 & 0 & 0 & 2/5 & 1/5 &
1/3 & 2/5 & 1/3\\ \hline 48&84 &1/5 & 1/3 & 1/3 & 2/5 & 2/5 & 1/5
& 1/5 & 1/5 & 1/3 & 1/2 & 1/2 & 2/3\\ \hline 49&84 &1/5 & 1/5 &
1/3 & 1/3 & 3/5 & 0 & 1/5 & 3/5 & 1/3 & 2/5 & 1/2 & 1/5\\ \hline
50&96 &1/5
& 2/5 & 2/5 & 2/5 & 1/2 & 1/5 & 0 & 3/5 & 1/5 & 2/5 & 3/5 & 1/3\\
\hline 51&96 &1/5
& 1/5 & 1/5 & 2/5 & 1/2 & 0 & 0 & 1/2 & 1/5 & 2/5 & 2/5 & 1/3\\
\hline 52&96 &2/5 & 2/5 & 2/5 & 2/5 & 3/5 & 0 & 0 & 3/5 & 2/5 &
2/5 & 4/5 & 1/3\\ \hline 53&96 &1/5 & 1/5 & 1/5 & 1/5 & 1/5 & 0 &
0 & 1/5 & 1/5 & 1/5 & 2/5 & 1/3\\ \hline 54&105 &1/3 & 1/3 & 1/3 &
2/5 & 3/5 & 1/5 & 1/5 & 2/5 & 2/5 & 1/2 & 1/2 & 2/5\\ \hline
55&105 &1/5 &
1/3 & 1/3 & 2/5 & 3/5 & 1/5 & 0 & 2/3 & 1/5 & 2/5 & 1/2 & 1/5\\
\hline 56&105 &1/5 & 1/5 & 1/3 & 1/3 & 2/5 & 1/5 & 1/5 & 1/5 & 2/5
& 1/3 & 1/2 & 1/2\\ \hline 57&105 &1/5 & 1/5 & 1/3 & 1/3 & 2/3 & 0
& 1/5 & 2/3 & 1/3 & 1/2 & 1/2 & 1/5\\ \hline 58&108 &1/3 & 2/5 &
2/5 & 2/5 & 2/5 & 1/5 & 0 & 1/3 & 1/3 & 2/5 & 2/3 & 3/5\\ \hline
59&108 &1/5 &
1/5 & 1/5 & 1/5 & 1/3 & 0 & 0 & 1/3 & 1/5 & 1/5 & 2/5 & 1/5\\
\hline 60&120 &1/5 & 1/3 & 1/3 & 1/3 & 3/5 & 1/5 & 1/5 & 1/2 & 1/3
& 1/2 & 1/2 & 1/3\\ \hline 61&144 &1/3 & 2/5 & 2/5 & 2/5 & 1/2 &
1/5 & 0 & 1/3 & 1/3 & 2/5 & 2/3 & 3/5\\ \hline 62&144 &1/5 & 1/3 &
2/5 & 2/5 & 1/2 & 1/5 & 1/5 & 1/3 & 1/3 & 1/2 & 1/2 & 1/2\\ \hline
63&144 &1/5 & 1/5 & 1/3 & 2/5 & 1/2 & 0 & 1/5 & 1/2 & 1/3 & 1/3 &
1/2 & 1/3\\ \hline 64&144 &1/5 & 1/5 & 1/5 & 1/3 & 1/2 & 0 & 0 &
1/2 & 1/5 & 1/3 & 2/5 & 1/5\\ \hline 65&144 &1/3 & 1/3 & 2/5 & 2/5
& 3/5 & 0 & 1/5 & 3/5 & 2/5 & 1/3 & 2/3 & 1/5\\ \hline 66&144 &1/5
&
1/5 & 1/5 & 1/3 & 1/3 & 0 & 0 & 1/3 & 1/5 & 1/3 & 2/5 & 2/5\\
\hline 67&200 &1/5 & 2/5 & 2/5 & 1/2 & 1/2 & 1/5 & 1/3 & 1/3 & 1/3
& 2/3 & 1/2 & 1/2\\ \hline 68&200 &1/5
& 1/5 & 2/5 & 1/2 & 1/2 & 0 & 1/3 & 1/2 & 2/5 & 1/3 & 1/2 & 2/5\\
\hline 69&205 &1/5 & 1/3 & 1/3 & 1/3 & 2/5 & 1/5 & 0 & 1/3 & 1/5 &
1/3 & 1/2 & 1/2\\ \hline 70&220 &1/3 & 1/3 & 1/3 & 2/5 & 2/5 & 0 &
0 & 2/5 & 1/3 & 2/5 & 2/3 & 1/2\\ \hline 71&220 &1/5 & 1/5 & 1/3 &
1/3 & 1/3 & 0 & 1/5 & 1/3 & 1/3 & 1/5 & 1/2 & 1/3\\ \hline 72&225
&1/3 &
1/3 & 1/3 & 1/3 & 2/5 & 1/5 & 1/5 & 1/5 & 2/5 & 2/5 & 1/2 & 1/2\\
\hline 73&225 &1/5 & 1/3 & 1/3 & 1/3 & 2/3 & 1/5 & 1/5 & 3/5 & 1/3
& 1/2 & 1/2 & 1/3\\ \hline 74&240 &2/5 & 2/5 & 2/5 & 1/2 & 1/2 & 0 & 1/3 & 1/2 & 2/5 &
1/5 & 3/5 & 1/3\\ \hline 75&240 &1/5 & 1/5 & 1/5 & 1/2 & 1/2 & 0 &
1/5 & 1/2 & 1/5 & 1/3 & 1/3 & 1/3\\ \hline 76&240 &1/3 & 1/3 & 2/5
& 2/5 & 1/2 & 1/5 & 1/5 & 1/3 & 1/3 & 1/3 & 1/2 & 2/5\\ \hline
77&240 &1/5 & 1/3 & 1/3 & 2/5 & 1/2 & 1/5 & 0 & 2/5 & 1/5 & 2/5 &
1/2 & 1/2\\ \hline 78&240 &1/5 & 1/5 & 1/3 & 1/3 & 1/2 & 0 & 1/5 &
1/2 & 1/3 & 1/3 & 1/2 & 1/5\\ \hline 79&252 &1/3 &
1/3 & 1/3 & 1/3 & 2/3 & 0 & 1/5 & 2/3 & 1/3 & 2/5 & 3/5 & 1/5\\
\hline 80&300 &1/3 & 1/3 & 1/3 & 1/3 & 3/5 & 0 & 0 & 3/5 & 1/3 &
1/3 & 2/3 & 1/5\\ \hline 81&300 &1/5 & 1/3 & 1/3 & 1/3 & 1/3 & 1/5
& 0 & 1/5 & 1/5 & 1/3 & 1/2 & 3/5\\ \hline 82&360 &1/3 & 2/5 & 2/5
& 1/2 & 1/2 & 1/5 & 0 & 1/3 & 1/3 & 1/2 & 2/3 & 2/3\\ \hline
83&360 &1/5 &
1/3 & 2/5 & 1/2 & 1/2 & 1/5 & 1/5 & 1/3 & 1/3 & 1/2 & 1/2 & 1/2\\
\hline 84&360 &1/5 & 1/5 & 1/3 & 1/2 & 1/2 & 0 & 1/5 & 1/2 & 1/3 &
2/5 & 1/2 & 2/5\\ \hline \ea\]

 \[\ba{|c|c||c|c|c|c|c||c|c|c|c|c|c|c|} \hline
\sharp & length &\theta_1 &\theta_2 &\theta_3 &\theta_4 &\theta_5 &\sigma_{12} &\sigma_{23} &\sigma_{34} &\sigma_{45} &\sigma_{51} &\sigma_{13} &\sigma_{24}\\
\hline 85&400 &1/3 & 1/3 & 1/3 & 2/5 & 1/2 & 0 &
0 & 1/2 & 1/3 & 2/5 & 2/3 & 2/5\\ \hline 86&400 &1/5 & 1/3 & 1/3 &
1/3 & 1/2 & 1/5 & 0 & 1/2 & 1/5 & 1/3 & 1/2 & 1/3\\ \hline 87&432
&1/3 &
1/3 & 1/3 & 1/3 & 1/3 & 0 & 1/5 & 1/3 & 1/3 & 1/5 & 3/5 & 1/3\\
\hline 88&480 &2/5 & 2/5 & 1/2 & 1/2 & 1/2 & 0 & 1/5 & 1/2 & 1/2 &
1/3 & 4/5 & 1/2\\ \hline 89&480 &1/5 & 1/5 & 1/2 & 1/2 & 1/2 & 0 &
1/3 & 1/2 & 1/2 & 2/5 & 2/3 & 1/2\\ \hline 90&576 &1/3 & 1/3 & 1/3
& 1/3 & 1/2 & 0 & 1/5 & 1/2 & 1/3 & 1/5 & 3/5 & 1/5\\ \hline
91&580 &1/5 & 2/5 & 1/2 & 1/2 & 1/2 & 1/5 & 1/5 & 2/5 & 1/3 & 3/5
& 2/3 & 1/2\\ \hline 92&600 &1/3 & 1/3 & 2/5 & 1/2 & 1/2 & 0 & 1/5
& 1/2 & 2/5 & 1/3 & 2/3 & 2/5\\ \hline 93&600 &1/5 & 1/3 & 1/3 &
1/2 & 1/2 & 1/5 & 0 & 1/3 & 1/5 & 1/2 & 1/2 & 2/3\\ \hline 94&900
&1/3 &
2/5 & 1/2 & 1/2 & 1/2 & 1/5 & 1/5 & 1/3 & 2/5 & 2/5 & 3/5 & 1/2\\
\hline 95&900 &1/5 & 1/3 & 1/2 & 1/2 & 1/2 & 1/5 & 1/5 & 1/3 & 2/5
& 1/2 & 2/3 & 3/5\\ \hline 96&936 &1/3 & 1/3 & 1/3 & 1/2 & 1/2 & 0
& 1/5 & 1/2 & 1/3 & 1/3 & 3/5 & 1/3\\ \hline 97&1200 &2/5 &
1/2 & 1/2 & 1/2 & 1/2 & 1/5 & 0 & 1/3 & 2/5 & 1/2 & 4/5 & 2/3\\
\hline 98&1200 &1/5 & 1/2 & 1/2 & 1/2 & 1/2 & 1/3 & 0 & 1/3 & 1/5
& 1/2 & 2/3 & 2/3\\ \hline 99&1440 &1/3 & 1/3 & 1/2 & 1/2 & 1/2 &
0 & 1/5 & 1/2 & 1/2 & 2/5 & 4/5 & 1/2\\ \hline 100&2160 &1/3 & 1/2
& 1/2 & 1/2 & 1/2 & 1/5 & 0 & 2/5 & 1/3 & 1/2 & 4/5 & 3/5\\ \hline
101&3072 &1/2 & 1/2 & 1/2 & 1/2 & 1/2 & 0 & 1/5 & 1/2 & 1/2 & 1/3
& 4/5 & 2/5\\ \hline \ea\]

We divide the orbits from this list into the following types:

$A$: Orbit 1. A dihedral orbit of length $4$ with arbitrary
parameters.

$B$: Orbit 2. A dihedral orbit with rational parameters. Its
length depends on parameters and can be arbitrarily large.

$C$: Orbits 3 and 4 with one parameter.

$D$: Orbits 9, 10, 11. Tetrahedral orbits.

$E$: Orbits 12-30. Octahedral orbits.

$F$: Orbits 31-101. Icosahedral orbits.

$G$: Orbits 5, 6, 7.

$H$: Orbit 8.

$N$: The orbits obtained from ${\mathcal M_F}^{(4)}$ by addition of the unit matrix. These orbits are omitted in this list.

Orbit $1$ is described in \cite{girand}.

Orbits 8, 11, 17, 19, 20, 23, 24, 25, 26, 27, 28, 29, 30, 35,
36, 37, 40, 46, 47, 52, 53, 55, 56, 58, 59, 65, 66, 67, 68, 69,
70, 71, 74, 75, 76, 78, 80, 81, 82, 83, 84, 88, 89, 91, 92, 93,
94, 95, 97, 98, 100, 101 (52 orbits in total) appear in
\cite{mazocco}.

Also, we found that orbits 42 and 52 from Table 2 in
\cite{mazocco} have lengths $432$ and $1440$ respectively. We
think that in \cite{mazocco} possibly there are misprints in
lengths of these two orbits because these lengths repeat lengths
in the adjacent rows in the table. If one fix it, these orbit
coincide with orbits 87 and 99 from Table 9 in the present paper,
respectively.

\section{Conjecture about classification of ${\mathcal M}^{(n)}$ for any $n$}

We conject that for any $n\geq 3$ there exist only the
following finite orbits (including cases $n = 4,\,5$ which are
already classified):

Type $I$. Orbits of {\it triangular} ${\mathcal M}^{(n)}$'s, where all matrices have a common eigenvector (not
considered in this paper; need a separate classification),

Types $A$, $B$, $D$, $E$, $F$. Orbits of these types exist for all $n$. All
matrices in these orbits belong to a subgroup of $SU(2)$, including:

$A$: belonging to the dihedral group, possibly infinite. Each
element of such orbit contains two matrices with zeros on the main
diagonal and $n-2$ diagonal matrices; there are $n-2$ arbitrary
parameters. Length of the orbit is $2^{n-3}$.

$B$: belonging to any finite dihedral group. There are $\lfloor n/2 \rfloor-1$
sub-types of such orbits, we denote them with integer number $m \in
[2,\lfloor n/2 \rfloor]$. Each element of an orbit of each sub-type contains $2
m$ matrices with zeros on the main diagonal, and $n-2 m$ diagonal
matrices; there are $n-2$ rational parameters. The length of the orbit
is by order of magnitude as large as the common denominator of all
parameters raised to the power $2 m - 2$.

$D$: tetrahedral orbits: all matrices belong to the tetrahedral
group.

$E$: octahedral orbits: all matrices belong to the octahedral
group.

$F$: icosahedral orbits: all matrices belong to the icosahedral
group.

Type $C$: this type exists only for $n=4,\,5,\,6$ and contains one parameter.
For $n = 6$ it has the form
\[
M_1 = \left(\ba{ll}X&0\\1&X^{-1}\ea\right),\quad M_2 =
\left(\ba{ll}X^{-1}&-1\\0&X\ea\right), \quad M_3 =
\left(\ba{ll}-X^{-1}&1\\0&-X\ea\right),\]
\be\label{sixc} M_4 = \left(\ba{ll}X&0\\1&X^{-1}\ea\right),\quad
M_5 = \left(\ba{ll}X^{-1}&-1\\0&X\ea\right), \quad M_6 =
\left(\ba{ll}X^{-1}&-1\\0&X\ea\right),\ee

and for $n=5,\,4$ reductions of this orbit.

Types $G$ and $H$ exist for $n = 4,5$. For $n=5$ there are exceptional orbits (\ref{noletter105}) and
(\ref{noletter192}), and for $n=4$ their reductions.

Type $K$: for $n=3$ arbitrary tuple of matrices.

Type $M$: For $n = 4$ in every orbit of types $A$, $B$, $C$, $D$,
$E$, $F$, $G$, $H$ we can replace $\{\theta_1,\,
\theta_2,\,\theta_3,\,\theta_4\}$ with
\[\left\{\theta_1-\delta,\,\theta_2-\delta,\,\theta_3-\delta,\,\theta_4-\delta\right\},\quad \delta = \frac{\theta_1+\theta_2+\theta_3+\theta_4}{2},\]
or with
\[\left\{\theta_1-\delta,\,\theta_2-\delta,\,\theta_3-\delta,\,\delta-\theta_4\right\},\quad \delta = \frac{\theta_1+\theta_2+\theta_3-\theta_4}{2}.\]

This type is completely considered in \cite{painleve_6}, and such
replacements of $\theta$'s match the Okamoto transformations.

Type $N$: for any orbit we can trivially increase $n$ by addition of the unit
matrix, or by addition of the minus unit matrix and multiplying any
other matrix by $-1$.

The reason why we believe in this conjecture is that for every
higher $n$ the diversity of finite orbits becomes lower. This is
because more and more conditions force orbits to be finite.
Moreover, each finite orbit for higher $n$ ($n\geq 6$) must have a
reduction -- a finite orbit for lower $n$ (e.g. $n = 5$). We checked this
conjecture with a non-exact computer search, but
without conclusive results.

\section{Discussion}

By this algorithm we can construct the list of ${\mathcal
M_F}^{(n)}$'s from a the list of ${\mathcal M_F}^{(n-1)}$'s. Thus,
starting from the list of ${\mathcal M_F}^{(4)}$'s, we can step by
step get lists for all $n$.

We classified the finite orbits of monodromies under braid group
actions for five branching points in the Fuchsian system for
$2\times 2$ matrices, and made a conjecture for such
classification for six and more branching points.

Some of the orbits of five monodromy matrices are listed in
Calligaris and Mazzocco paper \cite{mazocco}, and also in Diarra's
\cite{diarra}, Girand's \cite{girand} and Tsuda's \cite{tsuda},
but the full list turned out to be rather bigger. E.g. new type
orbits (5, 6, 7 in Table 9) were found.

The method of \cite{mazocco} is similar to the method presented in this paper.
The variables
$p_{a,b...}$ in \cite{mazocco} correspond to $2\,\cos\left(\pi\,\sigma_{a,b...}\right)$
in the present paper.

However, many orbits from Table 9 in the present paper are absent
in \cite{mazocco}. One reason may be that in \cite{mazocco} it is
declared that the authors used only the exceptional orbits from
\cite{painleve_6} for construction. However, the orbit number 9
from the present paper, which is considered in detail in sections
5, 6, construction of which needs not only the exceptional orbits
from \cite{painleve_6}, but also orbits with parameters, is absent
in \cite{mazocco}. Another reason may be a limitation of the
arithmetical method employed in \cite{mazocco}, which can handle
only explicit numbers in radical form, e.g. $\sqrt{2}$ and
$\left(1+\sqrt{5}\right)/2$. Therefore, the orbits 5, 6, 7 from
the present paper, for which $p$ values include seventh roots of
unity, which cannot be expressed in radicals, are also absent in
\cite{mazocco}.

The {\it triangular} cases where all monodromy matrices have a
common eigenvector and can be simultaneously put into lower
triangular form, were classified by Cousin and Moussard in
\cite{cousin}. We left these results outside of the present paper,
because the method used in this paper is not applicable to such
cases.

Although the proof uses the computer, it is an exact proof,
because the computer was tasked with searching through a finite
number of options using integer arithmetic (although too large for
a human to check in a reasonable amount of time). The program was
monitored for abnormal situations and no such situation ever
happened. In the end the procedure finished regularly after
exhausting of the possibility space. The proof of the conjecture
for six and more matrices must be analytical, but we expect it to
be not too complicated, because under our conjecture the list of
orbits for six and more matrices is rather uniform.

Each algebraic solution of the Garnier system corresponds a finite
orbit of ${\mathcal M}^{(n)}$. That's why we think that each orbit
of ${\mathcal M_F}^{(n)}$ with its tuple of exact values of
$\theta$'s may generate one algebraic solution of the Garnier
system, or a finite number of algebraic solutions corresponding to
the symmetry group of this ${\mathcal M_F}^{(n)}$. The solutions
for which one of $\theta$'s differs by 2 from a given solution,
can be obtained from it with an algebraic transformation,
analogous to B\"acklund transformations group, see \cite{noumi}.

\section*{Conclusions}

The finite monodromies of the Fuchsian system for five $2\times 2$
matrices have been classified, except for the case in which all
monodromy matrices have a common eigenvector.

The conjecture about such classification for six and more matrices
is formulated.


\pdfbookmark[1]{References}{ref}



\end{document}